\documentclass[twocolumn]{aastex63}
\usepackage{textcomp}
\usepackage{txfonts}
\usepackage{graphicx}

\newcommand{\Lya}{Ly$\alpha$}

\received{}
\revised{}
\accepted{11 January 2021}
\submitjournal{AJ}

\shorttitle{Ly$\alpha$, H$\alpha$, and \ion{He}{1} in GJ~9827\,b and d}
\shortauthors{Carleo et al.}
\graphicspath{{./}{figures/}}
\begin{document}

\title{A multi-wavelength look at the GJ~9827 system\\ No evidence of extended atmospheres in GJ~9827\,b and d from {\it HST} and CARMENES data}

\correspondingauthor{Ilaria Carleo}
\email{icarleo@wesleyan.edu}

\author[0000-0002-0810-3747]{Ilaria Carleo}
\affiliation{Astronomy Department and Van Vleck Observatory, Wesleyan University, Middletown, CT 06459, USA}
\affiliation{INAF -- Osservatorio Astronomico di Padova, Vicolo dell'Osservatorio 5, I-35122, Padova, Italy}

\author[0000-0002-1176-3391]{Allison Youngblood}
\affiliation{Laboratory for Atmospheric and Space Physics, University of Colorado, Boulder, CO 80303, USA}

\author[0000-0003-3786-3486]{Seth Redfield}
\affiliation{Astronomy Department and Van Vleck Observatory, Wesleyan University, Middletown, CT 06459, USA}

\author[0000-0002-2891-8222]{Nuria Casasayas Barris}
\affiliation{Instituto de Astrof\'\i sica de Canarias, C/\,V\'\i a L\'actea s/n, 38205 La Laguna, Spain}
\affiliation{Departamento de Astrof\'isica, Universidad de La Laguna, 38206 La Laguna, Spain}

\author[0000-0002-1242-5124]{Thomas R. Ayres}
\affiliation{Center for Astrophysics and Space Astronomy, University of Colorado, Boulder, CO 80309}

\author[0000-0001-5028-7941]{Hunter Vannier}
\affiliation{Astronomy Department and Van Vleck Observatory, Wesleyan University, Middletown, CT 06459, USA}

\author[0000-0003-4426-9530]{Luca Fossati}
\affiliation{Space Research Institute, Austrian Academy of Sciences, Schmiedlstrasse 6, A-8041 Graz, Austria}

\author[0000-0003-0987-1593]{Enric Palle}
\affiliation{Instituto de Astrof\'\i sica de Canarias, C/\,V\'\i a L\'actea s/n, 38205 La Laguna, Spain}
\affiliation{Departamento de Astrof\'isica, Universidad de La Laguna, 38206 La Laguna, Spain}

\author[0000-0002-4881-3620]{John~H.~Livingston}
\affiliation{Department of Astronomy, University of Tokyo, 7-3-1 Hongo, Bunkyo-ku, Tokyo 113-0033, Japan}

\author[0000-0001-5928-7251]{Antonino F. Lanza}
\affiliation{INAF - Osservatorio Astrofisico di Catania, Via S. Sofia 78, I-95123, Catania, Italy}

\author[0000-0002-8052-3893]{Prajwal Niraula}
\affiliation{Department of Earth, Atmospheric and Planetary Sciences, Massachusetts Institute of Technology, Cambridge, MA 02139}

%----CARMENES CO-AUTHORS-------------
% Nortmann, Chen, Nagel, Stangret
%Emails
%  guochen@pmo.ac.cn
%  nortmann.astro@gmail.com
%  mstangret@iac.es
%  evangelos.nagel@hs.uni-hamburg.de
%---------------------------------------

\author[0000-0001-5052-3473]{Juli\'an D. Alvarado-G\'omez}
\affiliation{Leibniz Institute for Astrophysics Potsdam An der Sternwarte 16, 14482 Potsdam, Germany}

\author[0000-0003-0740-5433]{Guo Chen}
\affiliation{Key Laboratory of Planetary Sciences, Purple Mountain Observatory, Chinese Academy of Sciences, Nanjing 210023, PR China}

\author[0000-0001-8627-9628]{Davide Gandolfi}
\affiliation{Dipartimento di Fisica, Universit\`a degli Studi di Torino, via Pietro Giuria 1, I-10125, Torino, Italy}

\author{Eike W. Guenther}
\affiliation{Th\"uringer Landessternwarte Tautenburg, Sternwarte 5, D-07778 Tautenburg, Germany}

\author[0000-0003-4446-3181]{Jeffrey L. Linsky}
\affiliation{JILA, University of Colorado and NIST, Boulder, CO 80309-0440, USA}

\author[0000-0002-4019-3631]{Evangelos Nagel}
\affiliation{Th\"uringer Landessternwarte Tautenburg, Sternwarte 5, D-07778 Tautenburg, Germany}
\affiliation{Hamburger Sternwarte, Gojenbergsweg 112, D-21029 Hamburg, Germany}

\author[0000-0001-8511-2981]{Norio Narita}
\affiliation{Department of Astronomy, University of Tokyo, 7-3-1 Hongo, Bunkyo-ku, Tokyo 113-0033, Japan}
\affiliation{Instituto de Astrof\'\i sica de Canarias, C/\,V\'\i a L\'actea s/n, 38205 La Laguna, Spain}
\affiliation{Astrobiology Center, NINS, 2-21-1 Osawa, Mitaka, Tokyo 181-8588, Japan}
\affiliation{National Astronomical Observatory of Japan, NINS, 2-21-1 Osawa, Mitaka, Tokyo 181-8588, Japan}
\affiliation{JST, PRESTO, 7-3-1 Hongo, Bunkyo-ku, Tokyo 113-0033, Japan}

\author[0000-0001-8419-8760]{Lisa Nortmann}
\affiliation{Instituto de Astrof\'\i sica de Canarias, C/\,V\'\i a L\'actea s/n, 38205 La Laguna, Spain}
\affiliation{Institut f\"ur Astrophysik, Friedrich-Hund-Platz 1, D-37077 G\"ottingen, Germany}

\author[0000-0002-7260-5821]{Evgenya L. Shkolnik}
\affiliation{School of Earth and Space Exploration Arizona State University, 781 S Terrace Road, Tempe, AZ 85281, USA}

\author[0000-0002-1812-8024]{Monika Stangret}
\affiliation{Instituto de Astrof\'\i sica de Canarias, C/\,V\'\i a L\'actea s/n, 38205 La Laguna, Spain}
\affiliation{Departamento de Astrof\'isica, Universidad de La Laguna, 38206 La Laguna, Spain}

%%TO order 

%\collaboration{1}{(AAS Journals Data Scientists collaboration)}

%\author{Butler Burton}
%\affiliation{Leiden University}
%\affiliation{AAS Journals Associate Editor-in-Chief}
%\nocollaboration{1}

%\altaffiliation{AASTeX v6+ programmer}

%\collaboration{1}{(LaTeX collaboration)}

%\affiliation{AAS Director of Publishing}
%\affiliation{American Astronomical Society \\
%1667 K Street NW, Suite 800 \\

%\author{Scott Chernoff}
%\affiliation{IOP Publishing, Washington, DC 20005}




\begin{abstract}

% This example manuscript is intended to serve as a tutorial and template for
% authors to use when writing their own AAS Journal articles. The manuscript
% includes a history of \aastex\ and documents the new features in the
% previous versions as well as the new features in version 6.3. This
% manuscript includes many figure and table examples to illustrate these new
% features.  Information on features not explicitly mentioned in the article
% can be viewed in the manuscript comments or more extensive online
% documentation. Authors are welcome replace the text, tables, figures, and
% bibliography with their own and submit the resulting manuscript to the AAS
% Journals peer review system.  The first lesson in the tutorial is to remind
% authors that the AAS Journals, the Astrophysical Journal (ApJ), the
% Astrophysical Journal Letters (ApJL), and Astronomical Journal (AJ), all
% have a 250 word limit for the abstract\footnote{Note that manuscripts 
% submitted to the new Research Notes of the American Astronomical Society 
% (RNAAS) do \textbf{not} have abstracts.}.  If you exceed this length the
% Editorial office will ask you to shorten it. This abstract has 180 words.

GJ~9827 is a bright star hosting a planetary system with three transiting planets. As a multi-planet system with planets that sprawl within the boundaries of the radius gap between terrestrial and gaseous planets, GJ~9827 is an optimal target to study the evolution of the atmospheres of close-in planets with a common evolutionary history and their dependence from stellar irradiation. 
Here, we report on the {\it Hubble Space Telescope} {\it (HST)} and CARMENES transit observations of GJ~9827 planets b and d. We performed a stellar and interstellar medium characterization from the ultraviolet {\it HST} spectra, obtaining fluxes for \Lya\ and MgII of $F$(\Lya) = (5.42$^{+0.96}_{-0.75}$) $\times$ 10$^{-13}$ erg cm$^{-2}$ s$^{-1}$ and $F({\rm MgII})$ = (5.64$\pm$ 0.24) $\times$ 10$^{-14}$ erg cm$^{-2}$ s$^{-1}$. We also investigated a possible absorption signature in \Lya\ in the atmosphere of GJ~9827\,b during a transit event from {\it HST} spectra, as well as H$\alpha$ and \ion{He}{1} signature for the atmosphere of GJ~9827\,b and d from CARMENES spectra. We found no evidence of an extended atmosphere in either of the planets. This result is also supported by our analytical estimations of mass-loss based on the measured radiation fields for all the three planets of this system, which led to a mass-loss rate of 0.4, 0.3, and 0.1 planetary masses per Gyr, for GJ~9827\,b, c, and d respectively. These values indicate that the planets could have lost their volatiles quickly in their evolution and probably do not retain an atmosphere at the current stage.
%{\color{red} This sentence is just a repeat of the earlier red text (marked in red b/c I need to update the numbers} Furthermore, we performed a stellar and interstellar medium characterization from the ultraviolet HST spectra, obtaining surface fluxes for \Lya\ and MgII of $F_S$(\Lya) = (2.32$^{+0.42}_{-0.33}$) $\times$ 10$^6$ erg cm$^{-2}$ s$^{-1}$ and $F_S(MgII)$ = (2.41$\pm$0.10) $\times$ 10$^{5}$ erg cm$^{-2}$ s$^{-1}$. 
\end{abstract}

\keywords{Exoplanet astronomy: Exoplanet systems --- High
resolution spectroscopy --- stars: activity --- 
ISM: abundances }


%%%%%%%%%%%%%%%%%%%%%%%%%%%%%%%%%%%%%%%%%%%%%%%%%%%%%%%%%%%%%%%%%%%%%%%%%%%%%%%%%

\section{Introduction} \label{sec:intro}

The \textit{Kepler} mission \citep{Boruckietal2010} discovered that planets between the size of Earth and Neptune are the most common type of exoplanets in our Galaxy (i.e., \citealt{Boruckietal2011,Batalhaetal2013,Roweetal2015}). Defined as planets with radii between 1 and 4 $R_{\oplus}$, they do not have any analogue in our Solar System. This makes them very captivating targets for the study of their formation and evolution history, as well as understanding their compositions, interior structures, and atmospheres. Moreover, terrestrial planets may be the most attractive targets for the search of biosignatures.

One of the most interesting and still unexplained characteristics of the sub-Neptune sized planet population is the gap in the radius distribution around 1.6 $R_{\oplus}$ found by \cite{Fultonetal2017}. Planet below this radius may be naked rocky cores, while those above this value have retained their atmospheres.
%Planets below this radius have a rocky core, while those above this value present a gaseous atmosphere.
A possible explanation for this gap suggests that gas-rich super Earths (mainly solid, rocky planets with a radius up to 1.5\,$R_{\oplus}$) will retain or lose their envelope depending on the level of irradiation from their host stars \citep{lopez2012, Owenetal2017, Loydetal2020}. It might be also possible that the mass-loss can be caused by the luminosity of the cooling planetary cores (core-powered mass-loss mechanism, \citealt{Guptaetal2019}). In this context, observing small planets is essential to better understand the role of photo-evaporation in their evolution. Moreover, observing multi-planet systems offer an extra benefit, since such systems presumably formed under the same initial conditions (i.e., same age, same flux evolution) and provide a unique opportunity to compare the compositions of planets with different sizes, as well as atmospheric characteristics at different incident flux. 

In this paper, we present the case of GJ~9827, a K6V star discovered to host three transiting planets in 1:3:5 commensurability by Kepler/K2 (\citealt{niraulaetal2017, rodriguezetal2018}). \cite{Teskeetal2018} showed with archival radial velocity (RV) observations from the \textit{Magellan} II Planet Finder Spectrograph (PFS) that planet b has a mass of $\sim$ 8 $\pm$ 2 $M_{\oplus}$, classifying it as one of the densest planets with an iron mass fraction of $\gtrsim$50\%; while planets c and d did not show strong constraints on their masses. \cite{Prieto2018} refined the planetary parameters through FIES, HARPS, and HARPS-N RVs, finding that planet b is less dense than suggested by \cite{Teskeetal2018}, with a mass of 3.74 $\pm$ 0.49 $M_{\oplus}$. \cite{Prieto2018} also calculated the incident fluxes for the three planets, pointing to a rocky composition for planets b and c, and a gaseous composition for planet d, which could possibly retain an extended atmosphere.

\cite{rice2019} combined data from \cite{niraulaetal2017}, \cite{Teskeetal2018}, \cite{Prieto2018}, and a new HARPS-N RV dataset, more precisely constraining the planetary masses for GJ~9827 b and d. A more recent analysis by \cite{kosiarek2020} refined the ephemerides from \textit{Spitzer} observations and, adding all the RVs from the previous work to their multi-year HIRES RV follow-up, more precisely constrained the parameters of this multi-planet system. The orbital periods, masses and radii of the three planets by \citep{kosiarek2020} are reported in Table \ref{tab:planets} and are the ones used in our analyses. It is worth to notice that the radii of the three planets span the above-mentioned radius gap at $\sim$1.6 $R_{\oplus}$, making this system even more appealing for uncovering how super-Earths form and evolve. 
%In fact, it is thought that sub-Neptunes can be transformed into super-Earths through the mass-loss driven by either the photoevaporation \citep[e.g.,][]{Owenetal2017}, or by the cooling of planet's core \citep[e.g.,][]{Guptaetal2019}.
%Straddling this rocky/gaseous planets divide, GJ9827's planets are primary targets to investigate these theories. }
Straddling this rocky/gaseous planets divide, GJ~9827 is ideal for studying the simultaneous evolution of planets at different orbital distances, having the stellar properties, including age, controlled.

%characterizing each of these planets might add an important piece to the formation picture of the system and in turn to the formation theories. 

%Kosiarek:
%Pb=1.2089765 +- 2.3e − 06 days
%Mb=4.87 +- 0.37 Me
%Rb=1.529+-0.058 Re
%Pc=3.648095 +- 2.4e − 05 days
%Mc=1.92 ± 0.49 Re
%Rc=1.201+-0.046 Me
%Pd= 6.20183 +- 1e − 05 days
%Md=3.42 +- 0.62 Me
%Rd= 1.955+-0.075Re

A key aspect for the assessment of long-term habitability in a planetary system is the atmospheric mass loss of planets due to the high-energy environment and stellar wind of its host star. Several spectral lines sensitive to extended atmospheres have provided unique measurements of mass loss, including Lyman-$\alpha$ \citep[e.g.,][]{vidal2003,kulowetal2014,ehrenreich2015}, H$\alpha$ \citep[e.g.,][]{jensen2012,cauley2015,casasayasbarris2018}, and \ion{He}{1} 10830 \AA\, \citep[e.g.,][]{spake2018,Salz2018He,Nortmann2018Science, Allartetal2018}. The far and extreme ultraviolet radiation from the star is capable of heating the upper portions of hot planet atmospheres, causing atoms to escape and possibly driving atmospheric mass loss \citep[e.g.,][]{murrayclay2009, france2016, Linskyetal2020}. This can have serious consequences for the evolution of the planet; if the mass-loss rate is high enough, the entire atmosphere can escape leaving behind a bare rocky core \citep[e.g.,][]{baraffe2005}. This
is especially relevant for planets with masses $<$ 0.1 $M_{\rm J}$ , i.e., the Neptune and super-Earth
regime \citep{owen2013}. Even in a less catastrophic case, the atmospheric composition
of the planet can be highly altered \citep{lopez2012}.

 The photochemistry of planetary atmospheres hosted by cool stars is controlled by the two strongest UV emission lines in the stellar spectrum, Lyman-$\alpha$ (\Lya) and Mg\,{\small II} \citep{madhusudhan2019}. They can drive significant water loss \citep[e.g.,][]{luger2015,miguel2015} and cause abiotic sources of biosignature gases \citep[e.g.,][]{tian2014,harman2015}. However, even for the nearest stars, absorption from the local interstellar medium (LISM) dramatically removes \Lya\ photons from the line of sight \citep{wood2005,youngblood2016}. The LISM is a rich and complex collection of clouds that leads to a unique absorption profile, often with more than one absorbing cloud \citep{redfield2004a,redfield2008}. To aid in the reconstruction of the intrinsic stellar \Lya\ flux, we observed and characterized the \ion{Mg}{2} stellar emission, which samples a similar level of the stellar chromosphere as \Lya, and yet is significantly less altered by LISM absorption. Given then high column density for HI ($\log N \sim 18$ cm$^{-2}$), $>$70\% of the intrinsic stellar chromospheric emission HI line can be absorbed by the LISM, and in some cases it can be $>$90\% \citep{wood2005}, whereas the lower abundances and column densities for MgII ($\log N \sim 12$ cm$^{-2}$), mean that $<$20\% of the intrinsic stellar chromospheric emission MgII line is absorbed \citep{redfield_2002}. Fitting the MgII lines provides not only a characterization of the intrinsic chromospheric line shape for \Lya, but also a fit to the LISM absorption profile. Although \cite{wood2005} used the line shape of MgII to inform the line shape of \Lya, the simultaneous fitting of the \Lya, MgII and ISM absorption that we perform in our analysis is novel.

We present here {\it HST} and CARMENES transit observations aimed at characterizing the atmosphere of GJ~9827\,b, in particular to evaluate its transit signature in different wavelength domains. The CARMENES analysis also includes data for GJ~9827\,d, the only planet of this system with previous atmospheric characterization: \cite{kasper2020} investigated the 10,830 $\AA$ HeI triplet  spectra, finding no absorption feature. In Section \ref{sec:obs} we describe the observations and data reduction for {\it HST} and CARMENES data. We also used {\it HST} data for characterizing the star, with \Lya, MgII, and XUV fluxes estimation in Section \ref{sec:stellar}. We investigate the possibility of an atmospheric planetary signal in Section \ref{sec:planet} and finally present our discussion in Section \ref{sec:discuss} and conclusion in Section \ref{sec:concl}. \\

\begin{deluxetable*}{cccc}
\tablenum{1}
\tablecaption{\label{tab:planets} Summary of GJ~9827 planets' parameters by \cite{kosiarek2020} adopted for our analyses.}
\tablewidth{0pt}
\tablehead{
\colhead{Planet} & \colhead{Orbital Period} & \colhead{Mass} & \colhead{Radius}   \\
\colhead{} & \colhead{(days)} & \colhead{(M$_{\oplus}$)} & \colhead{(R$_{\oplus}$)}  
}
%\decimalcolnumbers
\startdata
b  &  1.2089765 $\pm$ 2.3$\times$10$^{-6}$ &   4.87 $\pm$ 0.37  &  1.529 $\pm$ 0.058 \\
c  &  3.648095 $\pm$ 2.4$\times$10$^{-5}$ &   1.92 $\pm$ 0.49  & 1.201 $\pm$ 0.046 \\
d  &  6.20183 $\pm$ 1$\times$10$^{-5}$ &   3.42 $\pm$ 0.62  & 1.955 $\pm$ 0.075  \\
\enddata
\end{deluxetable*}

%Pb=1.2089765 +- 2.3e − 06 days
%Mb=4.87 +- 0.37 Me
%Rb=1.529+-0.058 Re
%Pc=3.648095 +- 2.4e − 05 days
%Mc=1.92 +- 0.49 Re
%Rc=1.201+-0.046 Me
%Pd= 6.20183 +- 1e − 05 days
%Md=3.42 +- 0.62 Me
%Rd= 1.955+-0.075Re

%%%%%%%%%%%%%%%%%%%%%%%%%%%%%%%%%%%%%%%%%%%%%%%%%%%%%%%%%%%%%%%%%%%%%%%%%%%%%%%%%

\section{Observations and data reduction} \label{sec:obs}
%%%%%%%%%%%%%%%%%%%%%%%%%%%%%%%%%%%%%%%%%%%%%%%%%%%%%%%%%%%%%%%%%%%%%%%%%%%%%%%%% 

\subsection{{\it HST} data}\label{sec:hstdata}
A four-orbit {\em HST}\/ pointing on GJ~9827 was carried out with the Space Telescope Imaging Spectrograph (STIS) during the 28 August 2018 transit of planet b as part of the Cycle 25 program GO-15434 (PI: S. Redfield).  The first-order far-ultraviolet (FUV) grating G140M was used, with its 1222~\AA\ setting, covering the wavelength interval 1194--1250~\AA, at a resolution of $R\equiv \lambda/\Delta\lambda\sim~10^4$. The chromospheric \ion{H}{1} 1215~\AA\ \Lya\ line is formed in the range 1-3$\times10^4$~K.
Other important emission lines in this region are the \ion{Si}{3} 1206~\AA\ resonance transition ($T\sim 5\times10^4$~K) and the \ion{N}{5} 1240~\AA\ doublet ($T\sim 2\times10^5$~K), although both of these were expected to be too faint to be significantly detected in the relatively brief FUV exposures of this faint, 10th-magnitude mid-K-type star. Nevertheless, the peak signal-to-noise (S/N) per 2-pixel resolution element (resel) in the combined spectrum at \Lya\ was about 35. The 52{\arcsec}$\times$0.1{\arcsec} narrow long slit was chosen to minimize geocoronal \Lya\ contamination from the upper atmosphere of the Earth. After the guide stars were acquired, a peak-up was performed to center GJ~9827, using the 31{\arcsec}$\times$0.5{\arcsec}~NDC long slit in the visible at low resolution with the STIS CCD.  The rest of the initial orbit was occupied by the first G140M exposure, of 1.8~ks.  The remaining three orbits had single G140M exposures, of 2.9~ks each.  A summary of the four FUV observations is provided in Table~\ref{tab:obs}.  In relation to the planetary transit, the first two exposures were pre-ingress, the third was in-transit, while the final was post-egress. 

Three months later, a single-orbit out-of-transit near-ultraviolet (NUV) spectrum of the \ion{Mg}{2} 2800~\AA\ region of GJ~9827 was taken, again with STIS, but now using the high-resolution echelle E230H with its setting 2713~\AA\ (2574--2851~\AA) and the ``spectroscopic'' slit 0.2{\arcsec}$\times$0.09{\arcsec} to achieve optimum resolution ($R{\sim}110,000$) for the narrow \ion{Mg}{2} absorptions. The 2713~\AA\ setting also captures an important \ion{Fe}{2} multiplet near 2600~\AA, although unfortunately the NUV continuum of the mid-K star was too weak for the \ion{Fe}{2} interstellar absorption to be detected (the peak S/N per resel at \ion{Mg}{2} 2796~\AA\ in the 1.8~ks exposure is 8).  A visible-light peak-up was performed with the same slit prior to the E230H echelle exposure.  The NUV observation is also described in Table~\ref{tab:obs}.

% \begin{table*}[htbp]
% \centering
% %\caption{}
% \caption{\label{tab:obs} Summary of {\em HST}/STIS exposures of GJ~9827. {\color{red} Use deluxetable for all tables}}
% \begin{tabular}{cccccc}
% \noalign{\smallskip}
% \hline
% \hline
% \noalign{\smallskip}
% Dataset  &  Mode-CENWAVE     &  Slit & UT Start Time & Exposure Time & Peak S/N [$\lambda$] \\
%               &      (\AA) & ({\arcsec}${\times}${\arcsec})  &  (yy--mm--dd.ddd) &  (ks) & (resel$^{-1}$ [\AA])\\
% \hline
% \hline
% \noalign{\smallskip}
% ODRL01010  &  G140M--1222 &   52${\times}$0.1  &  2018--08--23.548 &    1.79  & 15 [1216.0] \\
% \noalign{\smallskip}
% \hline
% \noalign{\smallskip}
% ODRL01020  &  G140M--1222 &   52${\times}$0.1  & 2018--08--23.602 &     2.89  & 18 [1216.0]   \\
% \noalign{\smallskip}
% \hline
% \noalign{\smallskip}
% ODRL01030  &  G140M--1222 &   52${\times}$0.1  & 2018--08--23.668 &     2.89  &  18 [1216.0]  \\
% \noalign{\smallskip}
% \hline
% \noalign{\smallskip}
% ODRL01040  &  G140M--1222 &   52${\times}$0.1  & 2018--08--23.734 &     2.89  &   18 [1216.0]  \\
% \noalign{\smallskip}
% \hline
% ODRL03010  &  E230H--2713 &   0.2${\times}$0.09  & 2018--12--01.266 &     1.82  &   8 [2796.5]   \\
% \noalign{\smallskip}
% \hline
% \end{tabular}
% \end{table*}

\begin{deluxetable*}{cccccc}
\tablenum{2}
\tablecaption{\label{tab:obs} Summary of {\em HST}/STIS exposures of GJ~9827.}
\tablewidth{0pt}
\tablehead{
\colhead{Dataset} & \colhead{Mode-CENWAVE} & \colhead{Slit} & \colhead{UT Start Time}  & \colhead{Exposure Time} & \colhead{Peak S/N [$\lambda$]} \\
\colhead{} & \colhead{(\AA)} & \colhead{({\arcsec}${\times}${\arcsec})} & \colhead{(yy--mm--dd.ddd)}  & \colhead{(ks)} & \colhead{(resel$^{-1}$ [\AA])}
}
%\decimalcolnumbers
\startdata
ODRL01010  &  G140M--1222 &   52${\times}$0.1  &  2018--08--23.548 &    1.79  & 15 [1216.0] \\
ODRL01020  &  G140M--1222 &   52${\times}$0.1  & 2018--08--23.602 &     2.89  & 18 [1216.0]   \\
ODRL01030  &  G140M--1222 &   52${\times}$0.1  & 2018--08--23.668 &     2.89  &  18 [1216.0]  \\
ODRL01040  &  G140M--1222 &   52${\times}$0.1  & 2018--08--23.734 &     2.89  &   18 [1216.0]  \\
ODRL03010  &  E230H--2713 &   0.2${\times}$0.09  & 2018--12--01.266 &     1.82  &   8 [2796.5]   \\
\enddata
\end{deluxetable*}

We performed reductions of the FUV G140M exposures
utilizing the pipeline-processed (CALSTIS) X2D files, which are wavelength- and flux-calibrated spatial/spectral images derived from rectified versions of the original long-slit stigmatic spectrograms.  The image $x$-direction is along the dispersion, with 0.053 \AA\ pixel$^{-1}$.  The image $y$-direction is the spatial (cross-dispersion) dimension, with 0.03\arcsec\ pixel$^{-1}$.  The image pixel flux densities tabulated in the X2D files, and associated photometric errors, are provided per \AA\ and per 0.0293\arcsec (the latter is the cross-dispersion angular pixel size), so the extracted spectrum (and photometric error) must be multiplied by that angular factor to yield flux densities (erg cm$^{-2}$ s$^{-1}$ \AA$^{-1}$).

\begin{figure}[]
   	\centering
	\includegraphics[width=1.25\linewidth, trim = 0cm 1cm 0cm 12cm]{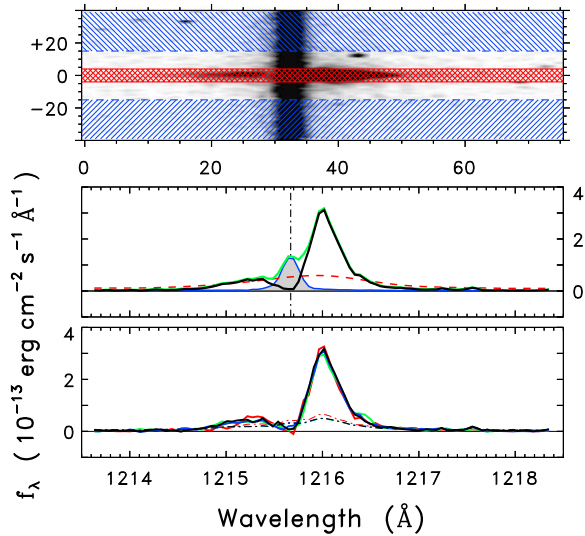}
	\caption{\textit{Top}: Co-added 2D spectral images (in pixels) of GJ~9827 G140M exposures. The red band represent flux extraction region for the stellar spectrum; the blue bands represent the region where the background, including geocoronal Ly-alpha, was sampled.  \textit{Middle}: Co-added 1D spectrum zoomed in the \Lya\ feature. The green curve is the total extracted spectrum; the gray shaded outlined by the blue curve is the background including the geocoronal emission; and black curve is the net flux (the total flux minus the background). The red dashed line is the photometric error.   \textit{Bottom}: Separate 1D spectra zoomed in the \Lya\ feature for the four exposures (ODRL01010 red, ODRL01020 green, ODRL01030 blue, ODRL01040 black). The thinner dotted-dashed curves are (smoothed) 3-sigma noise levels per resel for each of the spectra. The one for exposure 10 (in red) is higher than the others because the exposure was shorter.
	}
	\label{fig:redtom}
\end{figure}

The upper panel of Fig. \ref{fig:redtom} illustrates a co-added version of the 2D spatial/spectral image of the four G140M exposures.  The vertical extent of the image represents a ${\pm}$40 pixel slice in the detector $y$-direction ($\sim{\pm}$1.2\arcsec) centered on the apparent stellar \Lya\ feature.  The horizontal extent is 1200 pixels along the dispersion.  The narrow geocoronal stripe is conspicuous in the $y$-direction, bisecting the broader stellar \Lya\ feature.  The red band outlines a 9-pixel flux extraction region ($\sim$0.3\arcsec) for the stellar spectrum, while the blue dashed bands highlight flanking regions where the background was sampled. The two background bands are 25 pixels wide, beginning at $\pm$15 pixels from the center.  The wide bands increase the signal-to-noise for the background subtraction.  In practice, we eliminated the top three of the background values at each wavelength, in an effort to mitigate hot pixels.

%In practice, we applied a modified ``Olympic'' filter to the collection of background values at each wavelength, eliminating the top three, in an effort to mitigate hot pixels.  

The middle panel of Fig. \ref{fig:redtom} depicts the extracted 1D spectrum from the co-added image, zoomed into the \Lya\ feature.   The green tracing is the gross spectrum; blue with grey shadow is the background including the geocoronal \ion{H}{1} emission feature; and black is the net flux (gross--background).  The wavelength scale was set to place the geocoronal \Lya\ feature at its laboratory wavelength (1215.670~\AA).  The thin red dashed curve represents the 10\,$\sigma$ photometric error level (per resel), derived from the spatial/spectral values provided in the original X2D files, smoothed, for display, by a double pass of a rectangular filter 15 pixels wide.  The bottom panel shows the extracted \Lya\ features for the four exposures separately, where the geocoronal feature has been subtracted from the stellar profile.  There are small differences between the \Lya\ peaks of the four profiles, and the difference in the wing of the red profile (that corresponds to ODRL01010 of Table \ref{tab:obs}, the first observation of the sequence) around 1215 \AA\ is close to 3$\sigma$ in significance.

We reduced the single NUV exposure directly from the CALSTIS pipeline X1D file, which is a tabulation of extracted flux densities and associated photometric errors for 27 of the echelle orders contained in the original E230H-2713 spectral image.  We merged the individual orders together, tapering the overlapping zones to preserve the optimum S/N.  GJ~9827 is a relatively faint star for STIS high-resolution echelle spectroscopy, so the main features visible are the \ion{Mg}{2} 2803~\AA\ (``h'') and 2796~\AA\ (``k'') resonance doublet, the most important spectral signatures of the stellar chromosphere below the temperatures where the higher-excitation \Lya\ emission forms.  The peak S/N at the k line is about 8, and interstellar absorption is apparent in both emission cores.

The observed \Lya\ feature is the combination of the intrinsic stellar emission profile and the interstellar medium (ISM) attenuation profile \citep{wood2005}. The core of the stellar emission line originates in the lower transition region and upper chromosphere ($T\sim$~2--3$\times$10$^4$ K), while the outer wings form deeper in the stellar chromosphere.  The \Lya\ emission core is strongly attenuated by neutral hydrogen (\ion{H}{1}) and deuterium (\ion{D}{1}) gas over the 29.7~pc sightline to GJ~9827. The star's +31.9 km s$^{-1}$ radial velocity \citep{Prieto2018} shifts the stellar emission lines away from much of the ISM attenuation centered near 0 km s$^{-1}$ \citep{redfield2008}, giving the observed \Lya\ feature its asymmetric appearance. The \ion{Mg}{2} cores are less affected than \Lya\ owing to the much smaller cosmic abundance of magnesium.

%%%%%%%%%%%%%%%%%%%%%%%%%%%%%%%%%%%%%%%%%%%%%%%%%%%%%%%%%%%%%%%%%%%%%%%%%%%%%%%%%

\subsection{CARMENES data}
We observed one transit of GJ~9827b and one transit of GJ~9827d with the CARMENES spectrograph \citep[Calar Alto high-Resolution search for M dwarfs with Exo-earths with Near-infrared and optical Echelle Spectrographs;][]{CARMENES, CARMENES18} located at Calar Alto Observatory, in 13 August 2018 and 06 November 2018, respectively. CARMENES simultaneously covers the optical (VIS; $520$ to $960$ nm) and near-infrared range (NIR; $960$ to $1710$ nm), giving access to two important traces of planetary evaporation processes: the near-infrared \ion{He}{1} triplet at $10830~{\rm \AA}$ and the visible H$\alpha$ line at $6562.81~{\rm \AA}$. GJ~9827 is sufficiently bright (V= 10.1 mag, J= 7.98 mag) to be observed with both arms simultaneously. Both observations were performed in stare mode, taking continuous exposures before, during, and after the transit. The exposure times were set to be nearly the same for both arms, so that accounting for the different readout overheads of the VIS and NIR arms, the central time of each exposure was coincident in both. Resetting of the Atmospheric Dispersion Corrector took place between two exposures (i.e. during readout) and approximately every 30 minutes.

% \begin{table}[htbp]
% \centering
% \caption{\label{tab:obs_CARM} Summary of CARMENES observations.}
% \begin{tabular}{lcc}
% \noalign{\smallskip}
% \hline
% \noalign{\smallskip}
% Observing  night &  2018--08--13     & 2018--11--06\\
% \noalign{\smallskip}
% \hline
% \noalign{\smallskip}
% Planet transit &  GJ~9827 b &  GJ~9827 d  \\
% \noalign{\smallskip}
% \hline
% \noalign{\smallskip}
% Number of exposures  VIS &  43 &  40  \\
% \noalign{\smallskip}
% \hline
% \noalign{\smallskip}
% Number of exposures  NIR &  43 &  39  \\
% \noalign{\smallskip}
% \hline
% \noalign{\smallskip}
% Exposure time VIS (s)  & 200-192  &  200  \\
% \noalign{\smallskip}
% \hline
% \noalign{\smallskip}
% Exposure time NIR (s)  & 190-198  &  197  \\
% \noalign{\smallskip}
% \hline
% \noalign{\smallskip}
% Airmass change & 1.86-1.28  & 1.51-1.27-1.31   \\
% \noalign{\smallskip}
% \hline
% \noalign{\smallskip}
% Mean S/N in H$\alpha$ order  & 30  &  35  \\
% \noalign{\smallskip}
% \hline
% \noalign{\smallskip}
% Mean S/N in \ion{He}{1} order & 41 &  43  \\
% \noalign{\smallskip}
% \hline
% \end{tabular}
% \end{table}

\begin{deluxetable*}{lcc}
\tablenum{3}
\tablehead{}
\tablecaption{\label{tab:obs_CARM} Summary of CARMENES observations.}
\tablewidth{0pt}
\startdata
Observing  night &  2018--08--13     & 2018--11--06\\
\noalign{\smallskip}
\hline
\noalign{\smallskip}
Planet transit &  GJ~9827 b &  GJ~9827 d  \\
\noalign{\smallskip}
\hline
\noalign{\smallskip}
Number of exposures  VIS &  43 &  40  \\
\noalign{\smallskip}
\hline
\noalign{\smallskip}
Number of exposures  NIR &  43 &  39  \\
\noalign{\smallskip}
\hline
\noalign{\smallskip}
Exposure time VIS (s)  & 200-192  &  200  \\
\noalign{\smallskip}
\hline
\noalign{\smallskip}
Exposure time NIR (s)  & 190-198  &  197  \\
\noalign{\smallskip}
\hline
\noalign{\smallskip}
Airmass change & 1.86-1.28  & 1.51-1.27-1.31   \\
\noalign{\smallskip}
\hline
\noalign{\smallskip}
Mean S/N in H$\alpha$ order  & 30  &  35  \\
\noalign{\smallskip}
\hline
\noalign{\smallskip}
Mean S/N in \ion{He}{1} order & 44 &  47  \\
\enddata
\end{deluxetable*}

For the planet b observations, the exposure time at the beginning was $190$\,s but was then increased to $198$\,s, and the averaged S/N achieved is 44 in the \ion{He}{1} order and 30 in the H$\alpha$ order. On the other hand, for the planet d observations, the exposure times were $200$\,s and $197$\,s, with a S/N of 47 and 35 around \ion{He}{1} and H$\alpha$ orders, respectively. Due to a cloud crossing, the exposures taken at 19:22, 19:26, and 19:30 UT presented S/N below 30 in the \ion{He}{1} order and below 20 in the H$\alpha$ order. These exposures are discarded from the atmospheric analysis. In addition, for technical reasons, the observations were stopped from 19:59 to 20:16 UT. We note strong telluric contamination of the \ion{He}{1} region in both nights. A log table of the CARMENES observations is given in Table~\ref{tab:obs_CARM}.

We reduced CARMENES observations with the CARMENES pipeline CARACAL \citep[CARMENES Reduction And CALibration; ][]{CARACAL}, which performs bias, flat-relative optimal extraction \citep{CARMflatrelative}, cosmic-ray correction, and the wavelength calibration \citep{CARMcalibration}. The wavelengths are given in vacuum and the reduced spectra referenced to the terrestrial rest frame.\\

%%%%%%%%%%%%%%%%%%%%%%%%%%%%%%%%%%%%%%%%%%%%%%%%%%%%%%%%%%%%%%%%%%%%%%%%%%%%%%%%%

\section{Stellar and interstellar medium characterization} \label{sec:stellar}

UV stellar emission dramatically affects the fate of an exoplanet's atmosphere, both in the physical and chemical composition and through possible mass loss, especially when the planet closely orbits its host star. EUV and X-ray photons heat an exoplanet's upper atmospheric layers and drive mass loss, whereas UV photons are primarily responsible for photochemistry \citep{madhusudhan2019}. In particular, the \Lya\ emission line at 1215.67 \AA~dominates the far ultraviolet (FUV) spectrum of late-type stars and is the main source for the photodissociation of molecules such as water and methane. 
However, the \Lya\ emission line is heavily absorbed by neutral hydrogen in the interstellar medium (ISM) between the star and the Earth and is also contaminated by the geocoronal emission. This necessitates a reconstruction to recover the intrinsic stellar flux as seen by the planet's atmosphere. In the next sections we describe the \Lya\ and Mg II emission line reconstructions for GJ~9827, the derived properties of the line-of-sight ISM, and how GJ~9827's UV emission compares to other K dwarfs.

%%%%%%%%%%%%%%%%%%%%%%%%%%%%%%%%%%%%%%%%%%%%%%%%%%%%%%%%%%%%%%%%%%%%%%%%%%%%%%%%%
\subsection{Intrinsic Lyman-$\alpha$ and Mg II reconstruction} \label{sec:LyA_reconstruction}
To correct for the ISM attenuation of the \Lya~ and Mg II lines and recover the intrinsic stellar emission lines, we use the approach described in \cite{youngblood2016} and \cite{garciamunoz2020}, modified to jointly fit \Lya~and Mg II. This allows for stronger constraints on the physical parameters common to the two transitions. Using a Markov Chain Monte Carlo method \citep{ForemanMackey2013}, we simultaneously fit a model of the \Lya~and Mg II intrinsic stellar emission lines and ISM absorption lines to the STIS G140M and E230H spectra. %To better constrain the parameters of this model fit, we also simultaneously fit a similar model to the Mg II emission lines and ISM absorption in the STIS E230H spectrum, linking many of the related physical parameters. 

For \Lya, we parameterized the broad, intrinsic stellar emission line as a single Voigt profile in emission with a Gaussian in absorption for the line's self-reversal. For the narrow Mg II emission lines, we used a single Gaussian profile in emission for each line and a single Gaussian for each line's self-reversal. We required the radial velocities for the two Mg II lines to be identical and also required their FWHMs to be coupled such that FWHM$_k$ = 1.05$\times$FWHM$_h$, because high resolution stellar spectra indicate that the k line is $\sim$5\% broader than the h line (Brian E. Wood, private communication). We did not require the radial velocity of \Lya~to match the Mg II lines in case of small wavelength offsets between the two gratings. 

The ISM absorbers (H I, D I, Mg II) are modeled as single Voigt profiles in absorption (see Sec.~\ref{subsec:ISM} for justification for one component), each parameterized by a radial velocity $\varv$ (offset from their corresponding stellar emission line's radial velocity), Doppler width $b$, and species column density $N$. We assume that the ISM absorbers are coming from the same interstellar gas with the same kinematics, so all three ISM species were required to have the same radial velocity offset from their stellar emission line's radial velocity, and the Doppler widths were all connected assuming pure thermal broadening by $b_{MgII}$ = $b_{HI}$/$\sqrt{24.3}$ and $b_{DI}$ = $b_{HI}$/$\sqrt{2}$. While \ion{H}{1} and \ion{D}{1} are dominated by thermal broadening and thus this assumption should be fine, but heavier species like \ion{Mg}{2} will also have a significant contribution from turbulent broadening. Therefore, our value of $b_{MgII}$ will underestimate the true Doppler width of MgII. The D I and H I column densities were fixed to be $N$(D I)/$N$(H I) = 1.5$\times$10$^{-5}$ \citep{Wood2004}, and $N$(Mg II) was not linked to $N$(D I) or $N$(H I). The corresponding intrinsic emission lines and ISM absorption lines were multiplied together and convolved to the corresponding instrument resolution using the STIS G140M and E230H line spread functions.

%We explored variability in the intrinsic \Lya~line by performing the reconstruction on each of the four orbits' spectra individually. Fig.\ref{fig:bestfit4} shows some apparent variability, especially a lower amplitude during transit, but the 68\% confidence intervals overlap indicating this difference is not significant. 

%The resulting values for the intrinsic Ly-$\alpha$~flux are \textbf{I think I need to redo these on the data} erg cm$^{-2}$ s$^{-1}$ for the two pre-ingress spectra, the in-transit spectrum, and the post-egress spectrum, respectively. Fig. \ref{fig:boxplot} represents the posterior distributions of the four Ly-$\alpha$~fluxes with the median and 68\% and 95\% confidence intervals marked. 

Given the lack of an in-transit detection (see Section~\ref{sec:LyA_transit}), we co-added all four orbits' spectra to maximize the precision of the reconstruction. Figure~\ref{fig:LyA_bestfit} shows the best fit model and intrinsic profiles with 68\% and 95\% confidence intervals. We find an intrinsic \Lya~flux with 68\% confidence interval (5.42$^{+0.96}_{-0.75}) \times 10^{-13}$  erg cm$^{-2}$ s$^{-1}$. For Mg II, we find a best-fit integrated flux of the h line of (2.30$\pm0.16) \times 10^{-14}$ erg cm$^{-2}$ s$^{-1}$, and the integrated flux of the k line is (3.34$\pm0.17) \times 10^{-14}$ erg cm$^{-2}$ s$^{-1}$ (Figure~\ref{fig:mgii}). The total flux is F(Mg II h + k) = (5.64$\pm0.24) \times 10^{-14}$ erg cm$^{-2}$ s$^{-1}$.  These fluxes and the fitted ISM parameters (described in \ref{subsec:ISM}) are printed in Table \ref{tab:stellarism}.

\begin{figure}[ht]
   	\centering
	\includegraphics[width=1.0\linewidth]{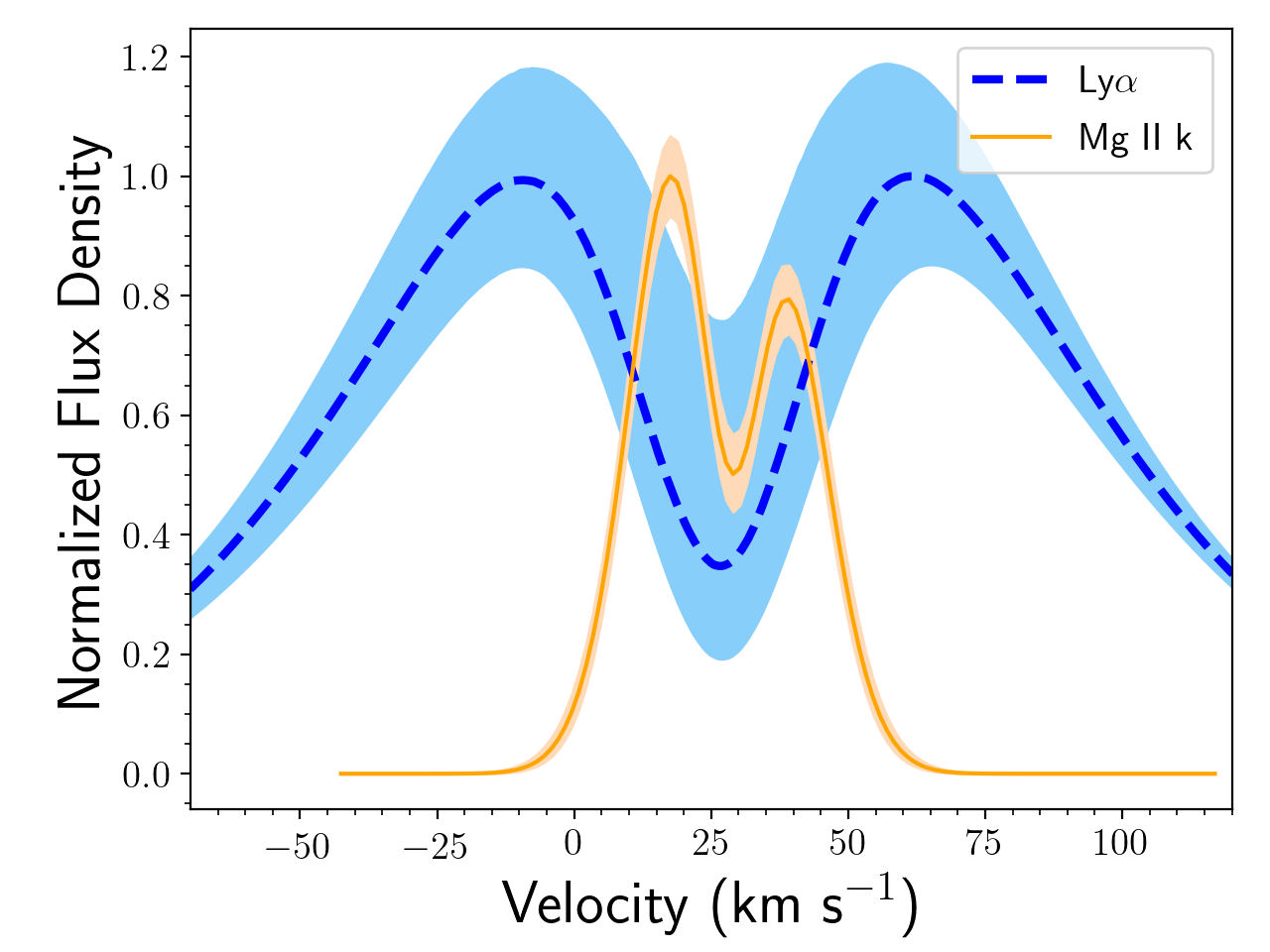}
	\caption{The intrinsic stellar profiles for \Lya~and Mg II are shown here, with ISM attenuation and instrumental resolution effects removed. Only the Mg II k line is shown for visual clarity, but the h line appears very similar to the k line. The profiles have been normalized to their peak values, and 68\% confidence intervals are shown as the shaded regions. }
	\label{fig:self_reversal_comparison}
\end{figure}

In Figure~\ref{fig:self_reversal_comparison}, we compare the line core shapes of the reconstructed \Lya~and Mg II lines, with ISM attenuation and instrument resolution effects removed. The \Lya~line is $>$100 km s$^{-1}$ broader than Mg II, as expected. Their self-reversal shapes are roughly similar, but there are significant discrepancies between the depth, width, and asymmetry of the two species. For \Lya, the self-reversal was only allowed to deviate $\pm$10 km s$^{-1}$ from line center due to significant degeneracy between the self-reversal depth and ISM absorption, but is centered almost exactly at line center. The self-reversal in the Mg II lines was allowed to vary more widely, and it is readily apparent from the observed spectra that the self-reversal is asymmetric (i.e., not centered exactly at the stellar velocity). The \Lya~self-reversal depth also appears larger than Mg II's, however, the uncertainty on \Lya's self-reversal parameters are extremely large as the line core is not observed due to severe ISM attenuation. %On the other hand, the Mg II self reversal parameters are constrained directly from the data as the ISM is sufficiently Doppler-shifted away from the narrow stellar emission line.

%In the next section, we describe the measured ISM absorbers.

\subsection{Interstellar Medium Characterization} \label{subsec:ISM}
We used the LISM Kinematic Calculator\footnote{\url{lism.wesleyan.edu/LISMdynamics.html}} \citep{redfield2008}, which calculates whether or not a cloud of LISM traverses any given sight line, in addition to the radial and traverse velocities of the clouds in a given direction. In the case of GJ~9827, the LISM Kinematic Calculator yields no traversing clouds along the sight line. We note that this is likely due to the boundaries of clouds not being well sampled by limited LISM datasets, and this sight line probably does traverse at least one LISM cloud. There are 5 clouds passing within 20$^{\circ}$ of GJ~9827's sight line, including the Local Interstellar Cloud (LIC). The radial velocities of these clouds range from --7 to +10 km s$^{-1}$ with a weighted average of --3.1$\pm$5.8 km s$^{-1}$. Given the limitations of our data (low S/N in Mg II and low spectral resolution for H I), we fit a single ISM cloud component to our spectra. This results in fitted parameters that are likely akin to an average of the true parameters of the multiple clouds along the sight line. From our reconstructions described in the previous section, we find for interstellar \ion{H}{1} the following parameters: $\varv_{HI}$ = --3.35$^{+2.81}_{-2.83}$ km s$^{-1}$, $\log_{10}$ $N($\ion{H}{1}$) = 18.22^{+0.08}_{-0.10}$ cm$^{-2}$, and $b_{HI}$ = 12.8$^{+2.2}_{-4.5}$ km s$^{-1}$. For interstellar \ion{Mg}{2}, we find log$_{10}$ $N$(\ion{Mg}{2}) = 12.11$^{+0.51}_{-0.59}$ cm$^{-2}$. Recall that \ion{Mg}{2}'s Doppler $b$ parameter is defined as $b_{HI}$/$\sqrt{24.3}$ and the offset between the ISM radial velocity and the intrinsic stellar radial velocity was defined to be the same for the two species. This gives $\varv_{MgII}$ = 2.45$^{+2.07}_{-2.31}$ km s$^{-1}$ and $b_{MgII}$ = 2.6$^{+0.4}_{-0.9}$ km s$^{-1}$. While we expect $b_{MgII}$ to be underestimated, this is within the reasonable range for the Doppler width of LISM \ion{Mg}{2} absorption \citep{redfield2004a}. Note that the absolute uncertainty in the STIS MAMA wavelength calibration is 0.5--1 pixels, or 6--12 km s$^{-1}$ for G140M at \Lya\ and 1.5--3.0 km s$^{-1}$ for E230H at Mg II.

%% LIC = 1.24 1.38
%% Mic = -6.70 1.27
%% Hyades = 9.11 1.40
%% Eri = 0.83 1.04
%% Cet = -6.33 0.57
%% weighted average = -3.1 \pm 5.8 km/s

The constraint on the \ion{Mg}{2} interstellar absorbers is weak because of the narrowness of the stellar emission line serving as a backlight for the interstellar Mg II ions to absorb against. The interstellar Mg II atoms are Doppler shifted almost --30 km s$^{-1}$ from an emission line with FWHM = 24 km s$^{-1}$, and no stellar continuum is detected around the stellar emission lines. 

%{\color{red} Hunter's description}

%(Hunter's original text) We also used the ISM fitting technique described in \cite{redfield_2002} to characterize the interstellar absorption feature. The low signal to noise (S/N) ratio of the data combined with the location of the absorption in the wing of the Mg II emission feature made it more difficult to distinguish from the intrinsic stellar spectra. To contend with these challenges, we first used the Kinematic Calculator from \cite{redfield2008} which provides the proximity of interstellar clouds to a given line of sight and corresponding cloud radial velocities. The velocity of a cloud (LIC) close to GJ9827 sight line corresponded to the velocity of the absorption we observed (~1 kms$^{-1}$). Further, because there should be symmetry in the wings of the stellar emission feature, we mirrored the profile which highlighted evidence of absorption. The low S/N ratio required us to force the column density values until we achieved a reasonable fit the absorption feature, resulting in a column density of $log_{10}$N(Mg II)=12.5, within 1-$\sigma$ error of the previous fitting results shown in Figure~\ref{fig:mgii}. We are confident this separate method provides further confirmation of interstellar absorption.

From our simultaneously-fitted \ion{H}{1} and \ion{Mg}{2} column densities, we calculate the ratio $N$(MgII)/$N$(HI) = 8.0$^{+18.3}_{-6.0} \times 10^{-7}$ for GJ~9827's sight line. This value overlaps with the $N$(MgII)/$N$(HI) ratio from \cite{Linsky2019} (($3.6^{+2.8}_{-1.6}) \times 10^{-6}$) at the 68\% confidence interval (2.0$\times$10$^{-7}$--2.6$\times$10$^{-6}$). 

\begin{figure}[ht]
   	\centering
	\includegraphics[width=1.0\linewidth]{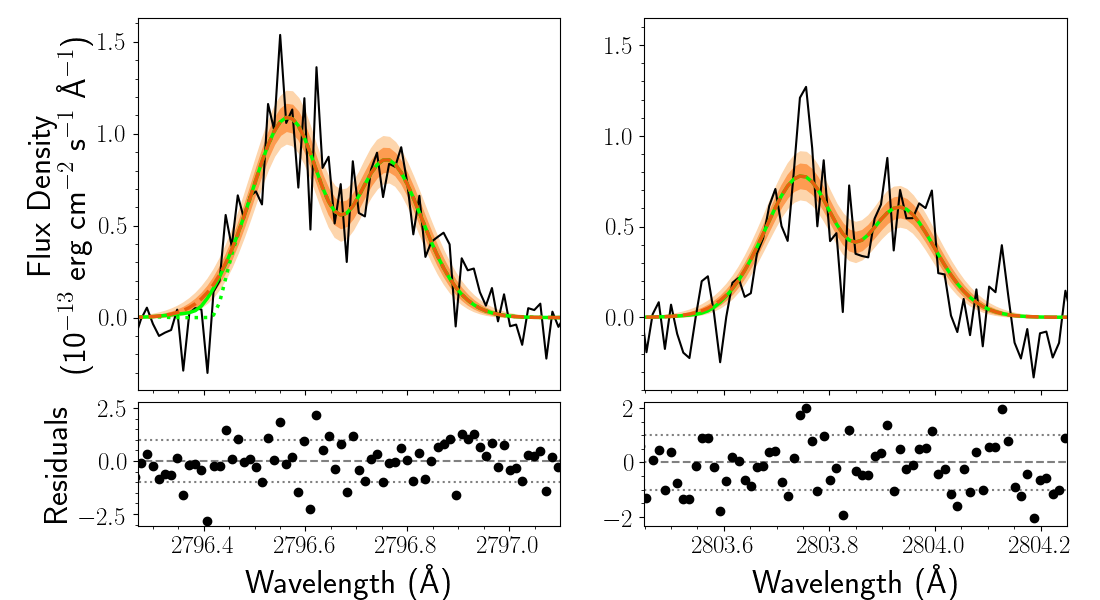}
	\caption{Reconstructed (orange dashed line) and fitted (attenuated; solid green line) Mg II profiles are shown. The ISM absorption is weak and centered at 2796.38 \AA~and 2803.56 \AA; the effect on the resultant profile is small. In the left panel, the dotted green line shows the model of the attenuated profile with the 2-$\sigma$ upper limit on the Mg II column density (log$_{10} N$(Mg II) $<$ 13.33). The bottom panels show the residuals (data-model)/(data uncertainty).}
	\label{fig:mgii}
\end{figure}

To verify our measurement of the \ion{Mg}{2} ISM absorption, we also applied the ISM fitting technique described in \cite{redfield_2002}. We assume the interstellar absorbers have a radial velocity equal to the LIC for this line-of-sight (+1.24 km s$^{-1}$; \citealt{redfield2008}), and we assume the wings of the \ion{Mg}{2} emission line to be symmetric. Masking the blueward, ISM-affected half of the line, we mirrored the redward half of the \ion{Mg}{2} k profile to create the assumed intrinsic stellar profile, which clearly indicated the presence of ISM absorption in the blue wing. The low S/N ratio in the wings of the line prevented a free fit to the column density. However, visual inspection using the mirrored profile led to a minimum column density log$_{10} N$(Mg II) $\approx$ 12.5. This value is 0.14 dex above the 1-$\sigma$ uncertainty range of the previous fitting results shown in Figure~\ref{fig:mgii}. Using the log$_{10} N$(Mg II)=12.5 value from our mirrored profile fitting of the Mg II line, we find $N$(MgII)/$N$(HI) = 1.9 $\times$ 10$^{-6}$, which is in agreement with the ratio value from \cite{Linsky2019}. The conclusion from using two different methods is that LISM absorption is present on the blueward wing, although it does not significantly alter the \ion{Mg}{2} emission profile. Our assumption of a single LISM absorption component and our choice of the stellar emission profile (i.e., automated Gaussian versus a mirrored profile) do not formally enter into our error analysis, and given the comparison to the LISM average $N$(MgII)/$N$(HI) ratio indicate that our \ion{Mg}{2} LISM column density may be slightly underestimated. 

We also searched for other ISM-affected spectral lines in the spectral range of STIS/E230H (2576 - 2823 $\AA$), such as the iron lines at 2586 and 2600 $\AA$, but the S/N is too low and the spectrum does not present any other spectral features. 

\begin{deluxetable*}{lcc}
\tablenum{4}
\tablehead{}
\tablecaption{\label{tab:stellarism} Stellar and ISM parameters.}
\tablewidth{0pt}
\startdata
Parameter  & Value & Units \\
\hline
\multicolumn{3}{l}{\bf \textit{Stellar fluxes}}  \\
\noalign{\smallskip}
Intrinsic \Lya\ flux F(\Lya)  & (5.42$^{+0.96}_{-0.75}$) $\times$ 10$^{-13}$ & \rm erg cm$^{-2}$ s$^{-1}$ \\
Intrinsic MgII $h$ flux $F$(MgII~h) & (2.30$\pm$ 0.16) $\times$ 10$^{-14}$ & \rm erg cm$^{-2}$ s$^{-1}$ \\
Intrinsic MgII $k$ flux $F$(MgII~k) & (3.34$\pm$ 0.17) $\times$ 10$^{-14}$ & \rm erg cm$^{-2}$ s$^{-1}$ \\
Intrinsic MgII $h+k$ flux $F$(MgII~h+k) & (5.64$\pm$ 0.24) $\times$ 10$^{-14}$ & \rm erg cm$^{-2}$ s$^{-1}$ \\
Surface \Lya\ flux $F_{S}$(\Lya) & (2.81$^{+0.50}_{-0.39}$) $\times$ 10$^6$ & \rm erg cm$^{-2}$ s$^{-1}$ \\
Surface MgII flux $F_{S}$(MgII) & (2.92$\pm$0.13) $\times$ 10$^{5}$ & \rm erg cm$^{-2}$ s$^{-1}$ \\
\hline
\multicolumn{3}{l}{\bf \textit{ISM absorbers' parameters}} \\  
HI radial velocity $\varv_{HI}$ & -3.4$^{+2.8}_{-2.8}$ & km s$^{-1}$ \\
HI Doppler width $b_{HI}$ & 12.8$^{+2.2}_{-4.5}$ & km s$^{-1}$\\
HI column density log$_{10} N$(HI) & 18.22$^{+0.08}_{-0.10}$ & cm$^{-2}$\\
MgII radial velocity $\varv_{\rm MgII}$ & 2.5$^{+2.1}_{-2.3}$ & km s$^{-1}$ \\
MgII Doppler width $b_{\rm MgII}$ & 2.6$^{+0.4}_{-0.9}$ & km s$^{-1}$\\
MgII column density log$_{10} N$(MgII) & 12.11$^{+0.51}_{-0.59}$ & cm$^{-2}$ \\
$N$(MgII)/$N$(HI) & 8.0$^{+18.3}_{-6.0}$ $\times$ 10$^{-7}$ & \\
\hline
\enddata
\tablecomments{The stellar radial velocity is +31.9 km s$^{-1}$ \citep{Prieto2018}. The $v_{HI}$ and $v_{MgII}$ values are derived from the same free parameter in the fit, a radial velocity offset from the stellar H I and Mg II emission lines centroids. The $v_{HI}$ and $v_{MgII}$ values differ because of differences in the absolute wavelength solution between the two gratings (G140M and E230H). The $b_{HI}$ and $b_{MgII}$ values are also derived from the same free parameter ($b_{HI}$) under the assumption of thermal line broadening ($b_{MgII}$ = $b_{HI}$/$\sqrt{24.3}$).}
\end{deluxetable*}

\subsection{GJ~9827 and other K dwarfs} \label{subsec:Kdwarf_comparison}

\begin{figure*}[ht]
   	\centering
	\includegraphics[width=1.0\linewidth]{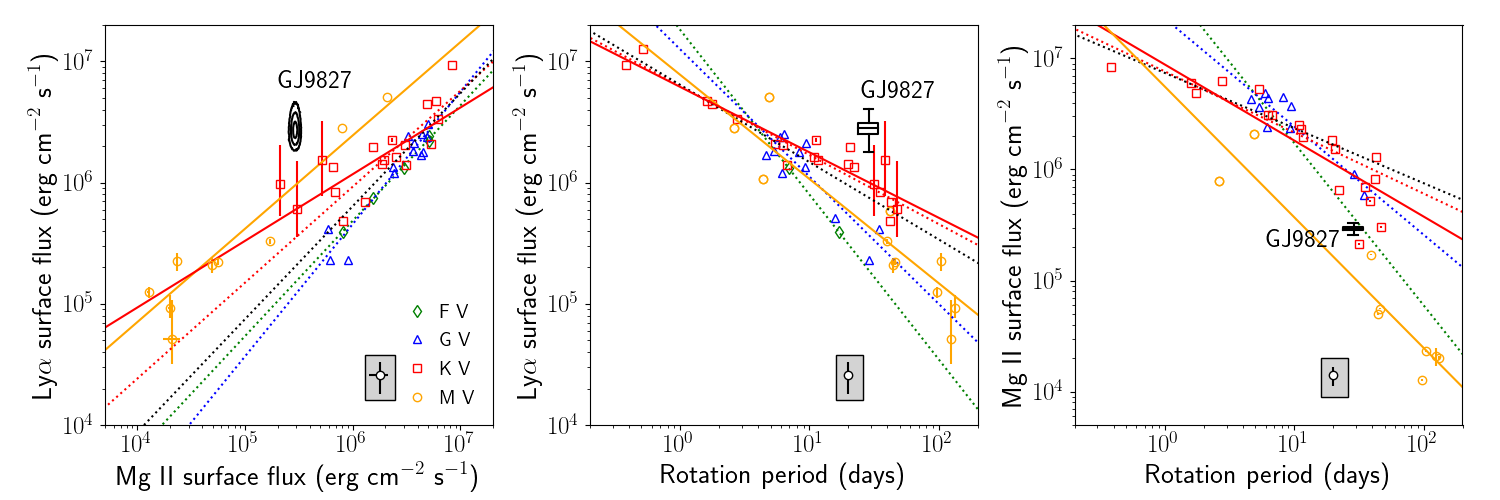}
	\caption{GJ~9827's \Lya~and Mg II surface fluxes (corrected for ISM absorption) and rotation period are compared to other F, G, K, and M stars from the literature \citep{wood2005,youngblood2016}. The literature sample is shown in colored, open symbols with corresponding best-fit linear regressions (dashed lines from \citealt{wood2005}, solid lines from \citealt{youngblood2016}). The example error bars in each panel apply to colored symbols without error bars. In the left panel, the 68\%, 95\%, and 99.7\% confidence intervals from our simultaneous \Lya\ and Mg II reconstruction are shown as black contours. In the right two panels, the black box and whisker symbol shows the median \Lya{} and Mg II surface flux with the 68\% and 95\% confidence intervals as the box and whiskers, respectively at the assumed stellar rotation period (28.72 days; \citealt{rice2019}).}
	\label{fig:Kdwarf_comparison}
\end{figure*}

We compare GJ~9827's \Lya~and Mg II fluxes to other K dwarfs with measured fluxes for both lines (Fig.~\ref{fig:Kdwarf_comparison}). To compare to data from \cite{wood2005} and \cite{youngblood2016}, we convert our fluxes into surface fluxes using the 0.579$\pm$0.018 R$_{\odot}$ radius from \citealt{kosiarek2020} and the 29.69 pc distance from Gaia DR2 to obtain $F_S$(\Lya) = (2.81$^{+0.50}_{-0.39}$) $\times$ 10$^6$ erg cm$^{-2}$ s$^{-1}$ and $F_S$(MgII) = (2.92$\pm$0.13) $\times$ 10$^{5}$ erg cm$^{-2}$ s$^{-1}$. GJ~9827's rotation period is poorly constrained, but appears to be between 15-30 days with a most likely value of 28.72 days \citep{rice2019}.

\begin{figure}[ht]
   	\centering
	\includegraphics[width=1.0\linewidth]{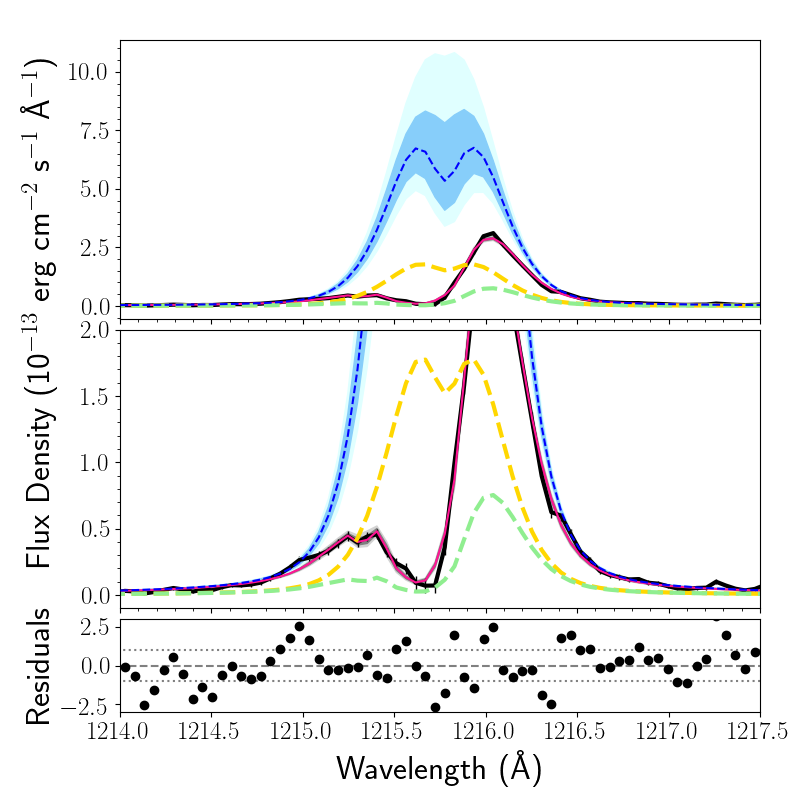}
	\caption{The reconstruction of the \Lya\ profile is shown in the top two panels (the middle panel is a zoomed version of the top panel with no changes). The STIS spectrum is represented in black with error bars (the error bars are generally smaller than the black line width). The best-fit model and 68\% and 95\% confidence intervals is shown as the pink line, dark gray shading, and light gray shading, respectively (the confidence intervals are also generally thinner than the width of the pink line). The intrinsic stellar emission line corresponding to the best-fit model is shown as the dashed blue line, with the 68\% and 95\% confidence intervals shown as dark-blue and light-blue shading, respectively. The bottom panel shows the residuals (data-model)/(data uncertainty). The dashed gold and green lines show how the intrinsic and observed (ISM attenuated) spectra would respectively appear if the intrinsic fluxes were consistent with Mg II - \Lya{} fluxes from the literature (Section~\ref{subsec:Kdwarf_comparison}). }
	\label{fig:LyA_bestfit}
\end{figure}

Compared to other K dwarfs of similar rotation period\footnote{The comparison K dwarfs with rotation period $>$15 days include HD 40307, HD 85512, HD 97658, $\alpha$~Cen B, 61 Cyg A, $\epsilon$~Ind, 40 Eri A, 36 Oph A, and $\sigma$~Gem. More information on all the K dwarfs in Figure~\ref{fig:Kdwarf_comparison} can be found in \cite{wood2005} and \cite{youngblood2016}.}, GJ~9827 has approximately 3.0 times less Mg II surface flux, and 2.8 times more \Lya\ surface flux. Mg II is commonly used as an estimator for the difficult-to-observe \Lya\ line \citep{wood2005}, and if we relied on the Mg II observation to estimate GJ~9827's \Lya\ emission, we would have underpredicted it by almost a factor of 5. Figure~\ref{fig:LyA_bestfit} shows how this MgII-derived \Lya\ profile (both intrinsic and observed) would appear and how the \Lya\ spectrum strongly rules out this flux level at the $>$3$\sigma$~level. This underestimation could have significant consequences for the atmospheres of GJ~9827's planets, because \Lya\ has a strong effect on photochemistry and as a proxy for the EUV \citep{Linsky2014}, could have implications for the atmospheric escape from the planets. 

 We estimate that there is a 0.001\% probability that GJ~9827's \Lya\ flux is consistent with the other K dwarfs in Figure~\ref{fig:Kdwarf_comparison}, and a 16\% probability that its Mg II flux is consistent with other K dwarfs. To calculate this, we drew 10$^6$ random samples from GJ~9827's \Lya\ and Mg II flux posterior distributions as well as 10$^6$ random samples from a normal distribution describing the red best-fit lines in the middle and right panels of Figure~\ref{fig:Kdwarf_comparison}, and determined what percentage of the posterior samples overlapped with samples from the best-fit lines. The normal distributions describing the best-fit lines had means equal to zero, and standard deviations equal to the standard deviations of all data points about the best-fit line, normalized by the best-fit line. For \Lya, the standard deviation is 0.29 and for Mg II it is 0.46. The samples from the \Lya\ and Mg II posterior distributions were also cast as differences from the best-fit line and then normalized by the best-fit line. %{\color{red} Note that there's a star near the GJ 9827 and describe it a little. Also note that 0.05\% is close to 3-sigma. Note there's more to these surface fluxes than rotation period and studying more objects like this can help us understand these objects.} 
 
What is causing the apparently significant discrepancy in the Mg II - \Lya\ flux ratio for this star? It is possible that this ratio is within the expected scatter of K dwarf UV flux-flux relations, and more UV spectroscopic observations K dwarfs are needed to quantify that typical scatter. Mg II and \Lya~form in slightly different regions in the stellar atmospheres and therefore their emission mechanisms are not exactly coupled, but in practice the scatter is likely dominated by the non-simultaneity of the Mg II and \Lya~observations. For example, the GJ 9827 Mg II and \Lya~observations were taken 100 days apart, or approximately 3.5 stellar rotation periods apart. Stellar surface inhomogeneities (e.g., active regions, faculae, plage) as well as evolution of these features responsible for much of the Mg II and \Lya~emission could cause deviations in the expected Mg II - \Lya\ flux ratio. 

Here we consider some additional effects, including metallicity and rotation evolution. GJ~9827 has slightly sub-solar metallicity ([M/H] = $-0.26$ to $-0.5$; \citealt{rice2019}), which could potentially explain a low Mg II flux relative to \Lya\ as well as GJ 9827's potentially anomalously-high \Lya\ flux. A detailed investigation into the effect of metallicity on relative Mg II and \Lya~line strengths in K dwarfs would be needed to determine that. Another possible explanation is the observed rotation evolution of \Lya\ luminosity compared to less optically thick chromospheric lines like C II.  Pineda et al. (\textit{under review}) showed with a sample of young and field age M dwarfs that \Lya\ luminosity declines much more slowly with increasing stellar rotation period (a proxy for stellar age) than other far-UV lines like C II. Assuming that Mg II behaves more similarly to lines like C II rather than \Lya, GJ~9827 could be at a point in its rotational evolution when its Mg II luminosity has decreased significantly but \Lya\ has not been impacted as much by stellar spindown. More UV observations of low-activity K dwarfs are needed to determine if this increasing \Lya/Mg II flux ratio with increasing rotation period is a real effect. 

%{\color{red} Could there be a circumstellar disk or something attenuating the Mg II?? I don't think so, the absorption would occur at the radial velocity of star... instead we see a pretty normal looking profile.}

%but $\epsilon$~Eri is also slightly sub-solar ([Fe/H] = -0.13 \cite{Santos2004}. Disentangling a metallicity effect from an activity effect is challenging.  This rotation period is intermediate to that of $\epsilon$~Eri (moderately active) and the inactive MUSCLES K dwarfs \citep{France2016}, but far lower than very active and/or young K dwarfs like Speedy Mic and PW And (both $<$2 days).

%%%%%%%%%%%%%%%%%%%%%%%%%%%%%%%%%%%%%%%%%%%%%%%%%%%%%%%%%%%%%%%%%%%%%%%%%%%%%%%%%
\section{Searching for planetary atmospheric absorption signal}\label{sec:planet}
\subsection{Investigating Lyman-$\alpha$} \label{sec:LyA_transit}
\begin{figure}[ht]
   	\centering
	\includegraphics[width=1.0\linewidth]{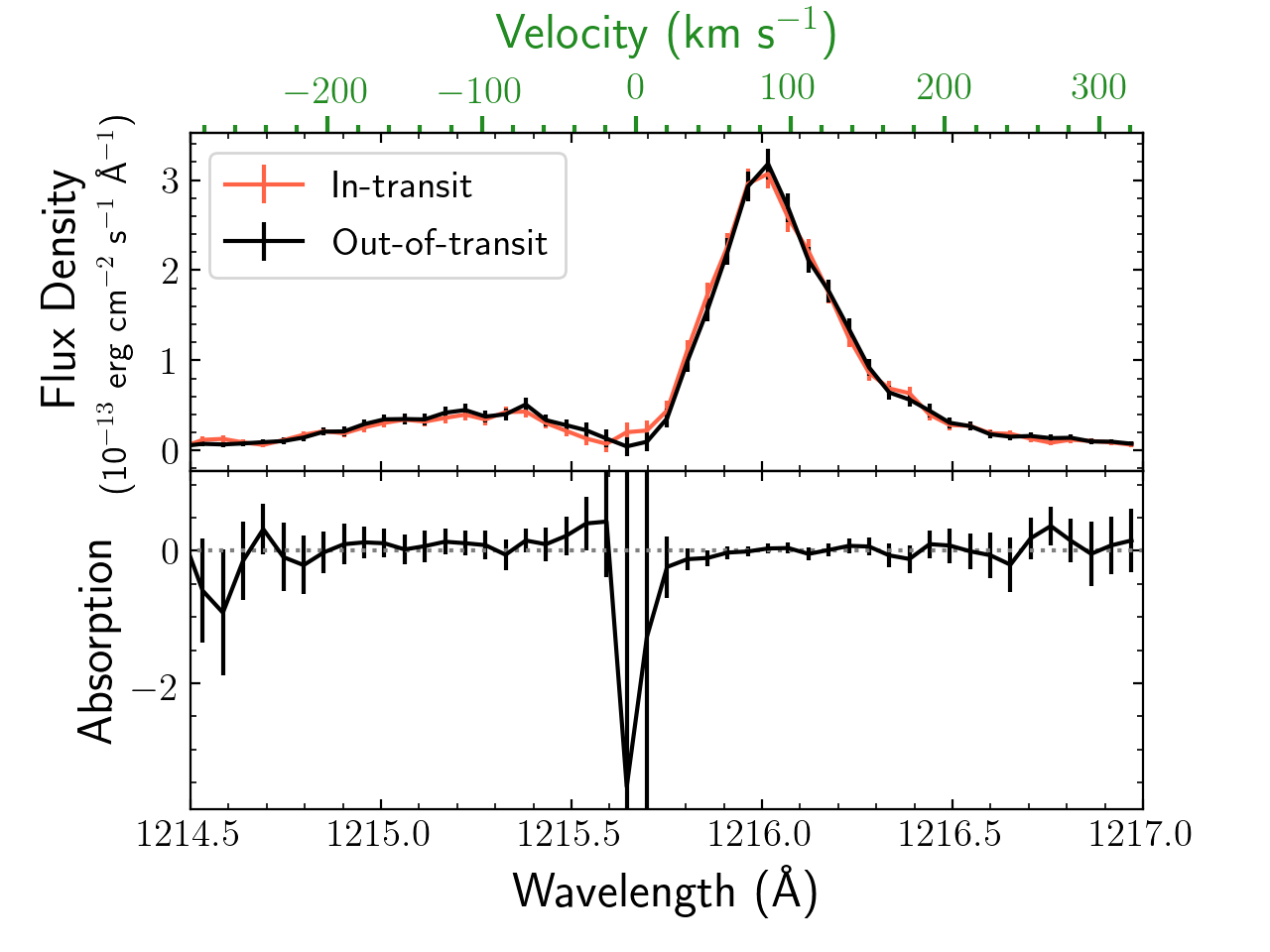}
	\caption{\textit{Top}. \Lya\ line of GJ~9827 during the transit (red line) and outside (black line). \textit{Bottom}. In-transit absorption depth, $1-F_{\rm IN}/F_{\rm OUT}$.}
	\label{fig:absdepth_tot}
\end{figure}

We investigate the behaviour of GJ~9827's observed \Lya\ profiles during the different phases of the transit. Starting from the four spectra obtained in Section \ref{sec:hstdata} (one per orbit) we calculated the in-transit absorption depth as $1-F_{\rm IN}/F_{\rm OUT}$, where $F_{\rm IN}$ is the flux of the \Lya\ line during the transit and $F_{\rm OUT}$ is the out-of-transit flux. 
Fig. \ref{fig:absdepth_tot} shows the out-of-transit (black solid line) and in-transit (red solid line) spectra for GJ~9827\,b, where the out-of-transit spectrum ($F_{OUT}$) is obtained by averaging the three out-of-transit spectra and the single in-transit spectrum represents $F_{IN}$. We find that the observed \Lya\ spectra are very similar in and out of the transit, and there is no evident planetary absorption during the transit. The largest apparent absorption depths occur in the spectral region most strongly contaminated by ISM absorption and geocoronal emission.

\begin{figure}[ht]
   	\centering
	\includegraphics[width=1.0\linewidth]{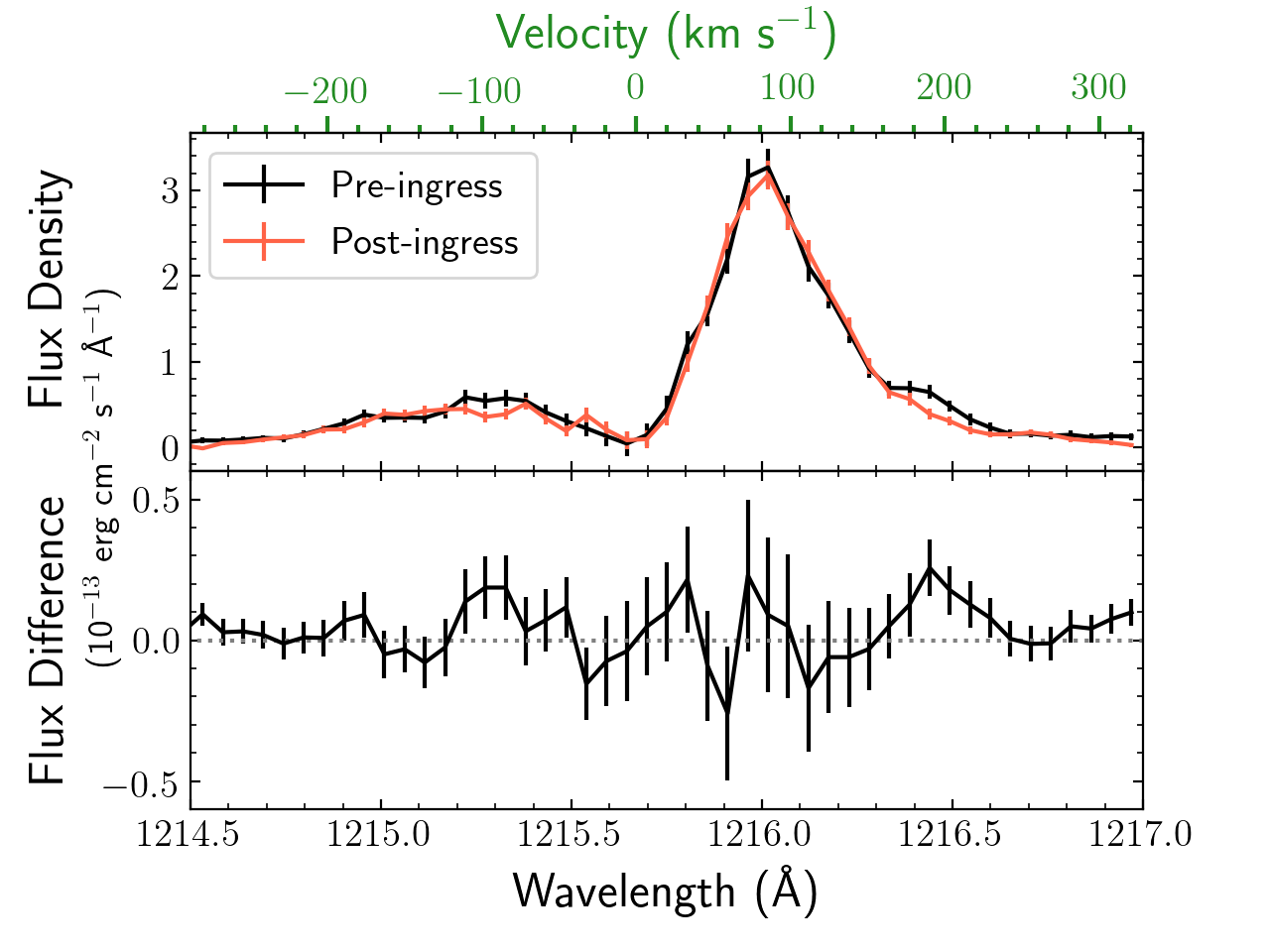}
	\caption{\textit{Top}. Pre-ingress Ly$\alpha$ spectrum (black line) (black line) and post-egress Ly$\alpha$ spectrum (red line). \textit{Bottom}. Difference between pre-ingress and post-egress spectra.}
	\label{fig:prepost}
\end{figure}

Furthermore, we compare pre-ingress and post-egress spectra, as in \cite{kulowetal2014}, to search for a possible atmospheric comet-like tail. \cite{McCannetal2019} showed that the stellar wind can significantly shape the planetary outflow, creating strong absorption signatures many hours before and after the optical transit. Fig. \ref{fig:prepost} shows pre-ingress and post-egress spectra in the top panel, and the difference between these spectra in the bottom panel. No evident difference is found. 

We integrated the flux in the \Lya~blue wing from -250 to -75 km s$^{-1}$ in the stellar rest frame (1214.787-1215.496 \AA) to obtain the average fluxes of each of the four orbits (units 10$^{-15}$ erg cm$^{-2}$ s$^{-1}$): 2.20$\pm$0.19, 2.28$\pm$0.15, 2.16$\pm$0.15, and 2.41$\pm$0.16. To obtain an upper limit of the size of the planet's H I atmosphere, we fit a transit model of an opaque sphere using a MCMC routine. We use the \texttt{batman} package \citep{Kreidberg2015} with transit ephemerides from Rice et al. 2019 and uniform limb darkening parameters. The size of the planet at \Lya\ relative to the star (R$_{Ly\alpha}$/R$_{\star}$) was the only free parameter, and we find 1-$\sigma$, 2-$\sigma$, and 3-$\sigma$ upper limits of 0.36, 0.48, and 0.57 for R$_{Ly\alpha}$/R$_{\star}$ in the blue wing. We repeated this upper limit calculation for the \Lya~red wing (+10 to +250 km s$^{-1}$ in the stellar rest frame; 1215.841-1216.813 \AA) and find average fluxes of each of the four orbits (units 10$^{-13}$ erg cm$^{-2}$ s$^{-1}$) of 1.10$\pm$0.04, 1.10$\pm$0.03, 1.09$\pm$0.03, and 1.13$\pm$0.03. The 1-$\sigma$, 2-$\sigma$, and 3-$\sigma$ upper limits on R$_{Ly\alpha}$/R$_{\star}$ in the red wing are 0.21, 0.27, and 0.32.

\subsection{Investigating \ion{He}{1} and H$\alpha$}
\label{sec:carm_analy}

\citet{Prieto2018} suggested that the GJ~9827 planetary system is an excellent laboratory to test atmospheric evolution and planetary mass-loss rates. We investigate the presence of evaporation traces in the atmospheres of GJ~9827\,b and GJ~9827\,d using CARMENES observations. In particular, we use the data obtained with the near-infrared channel to study the \ion{He}{1} triplet lines (at $10829.09~{\rm \AA}$, $10830.25~{\rm \AA}$ and $10830.34~{\rm \AA}$), and the visible channel data to study the H$\alpha$ line (at $6562.81~{\rm \AA}$).

\begin{figure*}[h]
   	\centering
	\includegraphics[width=0.98\linewidth]{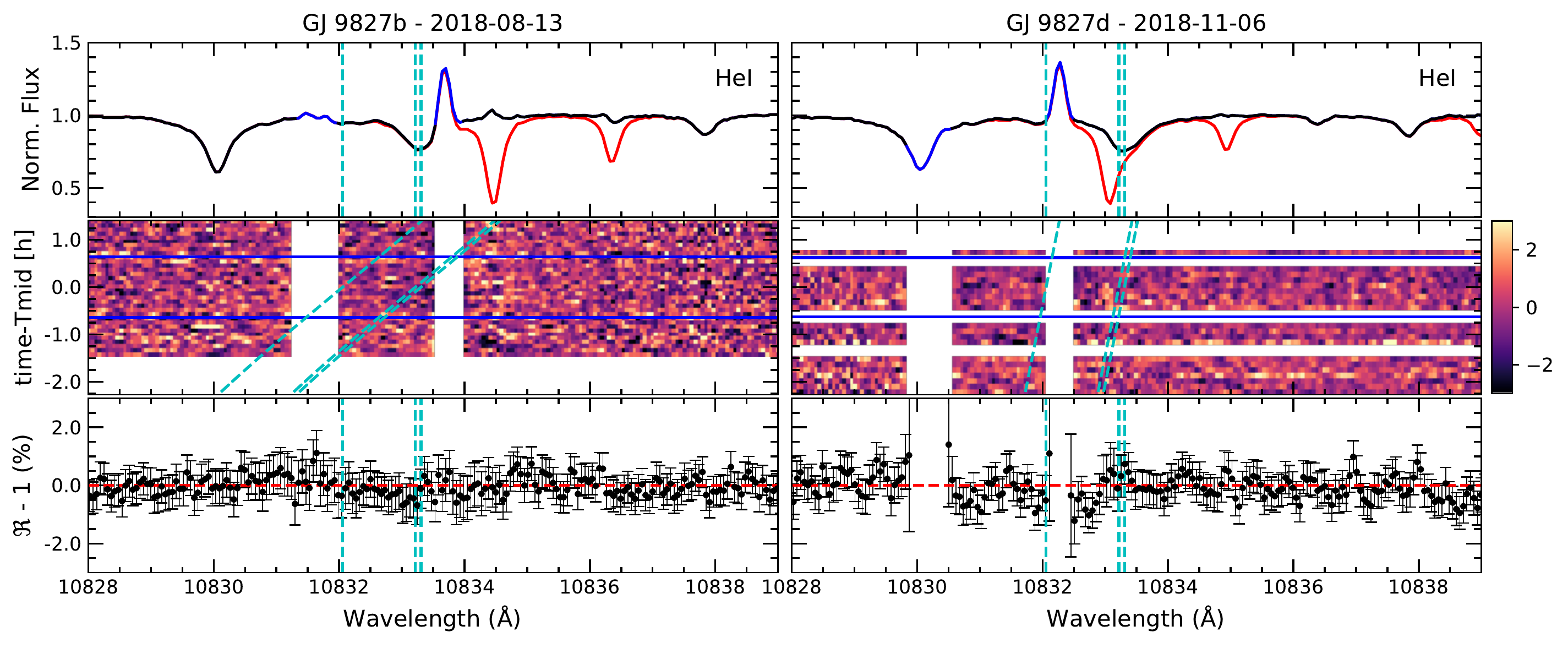}
	\caption{CARMENES transmission spectroscopy results around the near-infrared \ion{He}{1} triplet for GJ~9827\,b (left column) and GJ~9827\,d (right column). Top panel: GJ~9827's spectra in the \ion{He}{1} region. In the spectra we mark in red the telluric absorption lines and in blue the OH telluric emission lines region. In black we plot the spectra used to construct the transmission spectra, where the telluric absorption lines are corrected with Molecfit and the telluric OH emission lines are masked. The vertical/oblique cyan broken lines mark the positions of the triplet lines. Middle panel: 2D map of the residuals after dividing each spectrum by the master out-of-transit spectrum, in the stellar rest frame. The blue horizontal lines show the first and fourth contacts of the transit. The color scale shows the relative flux (F$_{in}$/F$_{out}$ -1) in \%. The white vertical regions correspond to the masked emission lines and the horizontal white regions to time stamps for which observations were discarded or missing. The vertical/oblique cyan broken lines mark the expected trace of the triplet lines during transit. The data are binned by 0.003 and 0.0008 in orbital phase for GJ~9827\,b and GJ~9827\,d, respectively. Bottom panel: The transmission spectrum for GJ~9827\,b and GJ~9827\,d. The red dashed line marks the zero absorption level. Again, the cyan dashed vertical lines show the position of the \ion{He}{1} triplet lines. All wavelengths are presented in the vacuum.}
	\label{fig:CARM_HeI}
\end{figure*}

\begin{figure*}[h]
   	\centering
	\includegraphics[width=0.98\linewidth]{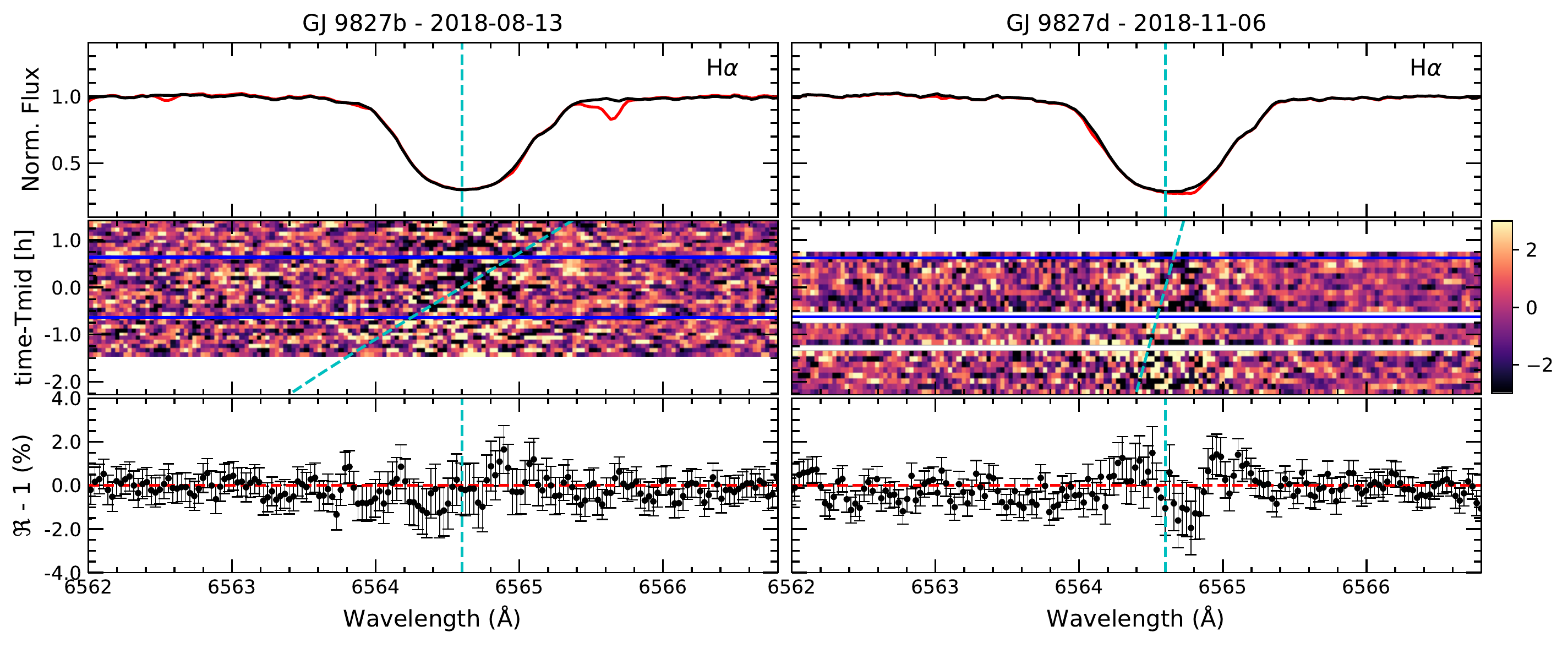}
	\caption{Same as Fig.~\ref{fig:CARM_HeI} but around the H$\alpha$ line. In this wavelength region only telluric absorption contamination is observed and no spectral regions are masked.}
	\label{fig:CARM_Ha}
\end{figure*}

As a first step, we correct the observed spectra of telluric absorption contamination using {\tt molecfit} (\citealt{Molecfit1} and \citealt{Molecfit2}), assuming the parameters presented in Nagel et al. (submitted) for the CARMENES instrumental line spread function model. In particular, the \ion{He}{1} region is contaminated by telluric absorption of water vapor and telluric emission of OH \citep{Nortmann2018Science, Salz2018He}. The water vapour absorption is corrected with {\tt molecfit} and the OH emission lines are masked, follwing the methods described in \citet{Palle2020}. The telluric line removal and masked regions are illustrated for each planet/night in the top panels of Figures~\ref{fig:CARM_HeI} and \ref{fig:CARM_Ha}.

After removing the telluric contamination, we can extract the transmission spectrum in both \ion{He}{1} and H$\alpha$ regions using the same approach, presented in different studies such as \citet{Wy2015}, \citet{Casasayas2019}, and \citet{Chen2020}. CARMENES observations are referenced to the Earth's rest frame. Thus, we shift the spectra to the stellar rest frame considering the barycentric radial velocity information and the system velocity ($31.95$\,km~s$^{-1}$; \citealt{Prieto2018}). After computing the ratio of each stellar spectrum to the master out-of-transit spectrum (combination of all out-of-transit spectra using the simple average) we move the residuals to the planet rest frame (see middle panels of Figures~\ref{fig:CARM_HeI} and \ref{fig:CARM_Ha}). To do this, we calculate the planet's radial velocity using the radial velocity semi-amplitudes $K_{p}^{\rm b} = 166.3$\,km~s$^{-1}$ and $K_{p}^{\rm d} = 98.2$\,km~s$^{-1}$, for GJ~9827\,b and GJ~9827\,d, respectively. These values are calculated assuming the stellar radial velocity semi-amplitude ($K_{\star}$), and the planet and star masses reported by \citet{kosiarek2020}, and using $K_p = K_{\star} M_{\star}/M_p$ \citep{Birkby2018}. Finally, we combine all in-transit residuals in the planet rest frame to obtain the transmission spectrum. In the \ion{He}{1} region, the masked intervals due to OH contamination are the same for all spectra in the stellar rest frame, but change when moving the spectra to the planet rest frame. When combining the in-transit residuals to extract the transmission spectrum, we only include the non-masked pixels in the calculation. The final transmission spectra are presented in the bottom panels of Figures~\ref{fig:CARM_HeI} and \ref{fig:CARM_Ha}. We note that different $K_{\star}$ and mass values are reported in the literature, which result in different $K_{p}$ values (see \citealt{rice2019} or \citealt{Prieto2018}, for example). However, these different $K_{p}$ values do not have significant impact on the derived transmission spectra.

It is important to notice, however, that using the parameters from \citet{kosiarek2020}, the mid-transit time of GJ~9827\,d's observations differs from that obtained using the parameters from \citealt{Prieto2018}, \citealt{rodriguezetal2018}, and \citealt{rice2019}. This difference is produced, mainly, by the differences in the derived orbital period value. The orbital period derived by \citet{kosiarek2020} differs by more than $30$\,s from the ones presented in the previous studies, and this difference is propagated along the different epochs. To be on the safe side, we have repeated the analysis using the parameters from the different references and the resulting transmission spectra do not show any significant feature in any of the cases.

In the two-dimensional residual maps presented in Figures~\ref{fig:CARM_HeI} and \ref{fig:CARM_Ha}, we are not able to visually distinguish absorption features during the transit that could have planetary origin, for either of \ion{He}{1} or H$\alpha$. The overall transmission spectrum does not show significant absorption features either. The excess absorption measured in the transmission spectra of GJ~9827\,b using a $0.7~{\AA}$ passband centred on the \ion{He}{1} and H$\alpha$ lines core is $-0.3\pm0.2~\%$ and $-0.2\pm0.2~\%$, respectively. On the other hand, for GJ~9827\,d, we measure $+0.2\pm0.2~\%$ and $-0.1\pm0.2~\%$ excess, respectively. The expected absorption signal ($\approx 2nH R_p/R_{\star}^2 $) of the annular area of one ($n=1$) atmospheric scale height ($H$) during the transit is around $3\times
10^{-5}$ for both planets, respectively, assuming an atmosphere dominated by H/He mixture and near solar composition ($\mu=2.3$). However, it has been observed in several exoplanet observations that the detected \ion{He}{1} signals are comparable to those created by an annular area of $n=100$ times the scale height (see \citealt{dosSantos2020}). Here, with only a single transit per planet, and the relatively small S/N ratio of the observations, especially in the lines core, we find no evidence for an extended H/He upper atmosphere around GJ~9827\,b or GJ~9827\,d. This is consistent with the non-detection \ion{He}{1} presented for GJ 9827 d by \citet{kasper2020}.

%The atmosphere of GJ~9827\,d has also been studied by \citet{kasper2020} using observations with the NIRSPEC instrument on the Keck II telescope. In consistency with the results presented here, this previous study concludes a non-detection of \ion{He}{1} in the sub-Neptune atmosphere.

%\textbf{More discussion?}
%%%%%%%%%%%%%%%%%%%%%%%%%%%%%%%%%%%%%%%%%%%%%%%%%%%%%%%%%%%%%%%%%%%%%%%%%%%%%%%%%

\section{Discussion} \label{sec:discuss}
The non-detection of planetary \Lya\ absorption during the transit of GJ~9827\,b can be the result of several factors. The particular radial velocity of the host star is such that the interstellar medium absorbs the line core and most of the blue wing of the \Lya\ line, while the red wing is almost intact. This is relevant, because past \Lya\ transit observations obtained for both hot Jupiters and warm Neptunes have shown that the planetary atmospheric absorption is strongest in the blue wing and it is typically caused by energetic neutral atoms, which are fast stellar wind protons that received electrons from slow planetary hydrogen atoms via charge exchange and are moving towards us at velocities of the order of 100\,km\,s$^{-1}$ \citep[e.g.,][]{vidal2003,kislyakova2014,ehrenreich2015,khodachenko2017,ildar2020}. Weak planetary atmospheric absorption in the red wing of the \Lya\ line, attributed to natural and thermal line broadening, has been observed before, for example in the case of HD209458b \citep{vidal2003}, but this absorption extends to a few tens of km\,s$^{-1}$ at maximum into both line wings \citep{kislyakova2014,khodachenko2017}. In the particular case of GJ~9827, the interstellar medium absorption is too close to the blue wing to enable detecting planetary absorption at low velocities and in the red wing the observed stellar emission flux may be too weak to allow detecting the planetary absorption, if present, above the noise level.

It has been suggested that for low-mass planets \Lya\ absorption may probe the presence of large amounts of water in planetary atmospheres; in this case the hydrogen is the result of water dissociation and further dragging of the lighter hydrogen in the upper layers as a result of the stellar high-energy irradiation \citep[e.g.,][]{bourrier2017}. In this case, the hydrogen originating from the atmospheric water vapor may be detectable, but, as described above, the specific configuration of the stellar emission and interstellar medium absorption hamper detecting the planetary absorption feature. Furthermore, a too large amount of water would also hamper the detection of hydrogen at \Lya\, because of the reduced atmospheric scale height due to the high mean molecular weight \citep{garciamunoz2020}.

It is also possible that absorption is not observed because the planetary atmosphere does not present enough hydrogen to be detectable. In fact, GJ~9827\,b has a bulk density of 7.47$^{+1.1}_{0.95}$ g cm$^{-3}$, thus consistent with a primarily rocky composition \citep{kosiarek2020}. Such a high average density may exclude the possibility that the planet hosts a primary, hydrogen-dominated atmosphere or an atmosphere holding large quantities of water. This is indeed the most likely explanation for the lack of planetary \Lya\ absorption. Given the high average density of the planet, the rather old age of the system, and the short orbital separation, we can expect that the planet has lost its primary, hydrogen-dominated envelope through escape in the first few hundreds of Myrs \citep[e.g.,][]{kubyshkina2018a,kubyshkina2018b}, developing then a secondary (e.g., CO$_{2}$-dominated) atmosphere as a result of magma ocean solidification. If this process happened while the star was still active, it is even possible that the planet has also lost this secondary atmosphere through hydrodynamic escape \citep{kulikov2006,tian2009,garciamunoz2020} leaving behind the bare surface exposed to the action of the stellar wind, which may have then led to the formation of a mineral exosphere \citep[e.g.,][]{miguel2011,vidotto2018}, not dissimilar from that of Mercury \citep[e.g.,][]{pfleger2015}.

The conclusion that GJ~9827\,b has most likely lost its primary hydrogen-dominated envelope is supported also by calculations of the expected planetary mass-loss rate. We employed the stellar distance and relations of \citet{Linsky2014} to estimate the stellar high-energy emission (X-ray and EUV; hereafter XUV) from the reconstructed \Lya\ flux, obtaining an XUV flux at 1\,AU of 13.01\,erg\,cm$^{-2}$\,s$^{-1}$. We further inserted the XUV flux, scaled to the distance of planet b, and the system parameters given by \citet{kosiarek2020} in the ``hydro-based approximation'' presented by \citet{kubyshkina2018c} that enables one to analytically derive hydrogen atmospheric mass-loss rates for planets below 40\,$M_{\oplus}$ accounting for all effects included in the hydrodynamic modelling (both core-powered mass-loss and photoevaporation). For GJ~9827\,b, we obtained a mass-loss rate of 3.6$\times$10$^{11}$\,g\,s$^{-1}$, that is about 1.9\,$M_{\oplus}$\,Gyr$^{-1}$ (or about 0.4 planetary masses per Gyr). Considering that the star is $\approx$10\,Gyr old \citep[certainly older than 5\,Gyr;][]{rice2019} and that the star was more active in the past, it is safe to conclude that the planet has lost its primary hydrogen-dominated atmosphere. This is further supported by the rather small restricted Jeans escape parameter $\Lambda$ \citep{fossati2017} of about 20.6, which alone indicates that the planet is subject to intense mass loss, partially driven by the high atmospheric temperature and low planetary gravity (i.e., core powered mass loss).

We followed the same procedure to estimate the atmospheric hydrogen mass-loss rates of GJ~9827\,c and GJ~9827\,d obtaining 1.0$\times$10$^{11}$\,g\,s$^{-1}$ and 5.1$\times$10$^{10}$\,g\,s$^{-1}$, respectively. We remark that for all three planets, the mass-loss rates obtained from the hydro-based approximation are within a factor of five from those obtained from directly interpolating the grid of hydrodynamic upper atmosphere models presented by \citet{kubyshkina2018b}. These values correspond to about 0.5\,$M_{\oplus}$\,Gyr$^{-1}$ (or about 0.3 planetary masses per Gyr) for GJ~9827\,c and about 0.3\,$M_{\oplus}$\,Gyr$^{-1}$ (or 0.1 planetary masses per Gyr) for GJ~9827\,d. These values and the bulk density of GJ~9827\,c (6.1\,g\,cm$^{-3}$) indicate that it is very unlikely the planet still holds part of its primary hydrogen-dominated atmosphere. It is therefore possible that GJ~9827\,c has followed an evolutionary path similar to that of GJ~9827\,b. 

For GJ~9827\,d, the lower bulk density of 2.51 g cm$^{-3}$ suggests that the planet may still host part of its primary atmosphere. This may be possible given that the planet appears to have a more sustainable mass-loss rate, suggesting that for this planet mass loss is primarily driven by atmospheric heating due to absorption of the stellar XUV emission, which is in general weaker than core-powered mass loss. However, the large difference between the mass-loss rates we obtained assuming a pure hydrogen atmosphere and those obtained from the constraint given by the non-detection of neutral He in the planetary transmission spectrum \citep[$<4.2\times10^8$\,g\,s$^{-1}$;][]{kasper2020} suggest that the planetary atmosphere may not be hydrogen-dominated. For GJ~9827\,d, mass loss is driven by the stellar high-energy emission and therefore its estimate directly depend on it. However a ten times lower stellar XUV emission compared to what is derived from the Ly$\alpha$ flux would bring the planetary mass-loss rate closer to the upper limit given by \citet{kasper2020}. Such lower stellar XUV emission would be close to what is indicated by the Mg II\,h\&k resonance lines. In fact, following \citet{linsky2013,Linsky2014}, the measured Mg II\,h\&k emission flux would imply an XUV flux at 1 AU of 5.97\,erg\,cm$^{-2}$\,s$^{-1}$ that is almost 3 times lower than what predicted from the Ly$\alpha$ emission flux. Furthermore one has to consider the rather large uncertainties on the employed conversions that make this value consistent with a ten times lower XUV flux. Nevertheless, even a ten times lower XUV flux would still be about ten times higher than what suggested by the non-detection of the He lines.

Interestingly, GJ~9827\,d has a bulk density similar to that of $\pi$\,Men\,c, which also has a predicted hydrogen mass-loss rate of the order of 10$^{10}$\,g\,s$^{-1}$ \citep{gandolfi2018,garciamunoz2020,ildar2020} and for which \Lya\ transit observations led to a non-detection of the planetary atmosphere \citep{garciamunoz2020}. %The hydrogen mass-loss rate obtained for GJ~9827\,c is particularly extreme, which is driven mostly by the low gravity (i.e., low planetary mass) and high temperature rather than by atmospheric heating due to absorption of the stellar XUV flux \citep{fossati2017,kubyshkina2018b}. However, the radial velocity measurements of \citet{rice2019} led only to an upper limit on the planetary mass, which brought \citet{rice2019} to suggest that the non-detection of the planet in the radial velocity data may have led to obtain a biased low planetary mass value. This would solve the contradiction between the low planetary bulk density and extremely large mass-loss rate. This is, however, not the case for GJ~9827\,d, which has well measured mass and radius. 
Transit observations of GJ~9827\,d, and of other similar planets such as $\pi$\,Men\,c, aiming at characterising their atmospheres would be very valuable for understanding the nature of puffy super-Earths \citep{garciamunoz2020}.\\

\section{Conclusions \label{sec:concl}}
In this paper we presented a search for exospheres around two planets orbiting GJ~9827, a K6 bright star discovered to host three super-Earths in 1:3:5 commensurability from the Kepler/K2 mission. We observed GJ~9827\,b with {\it HST} and GJ~9827\,b and d with CARMENES during transit in order to characterize their atmospheres via the \Lya, H$\alpha$, and He I transitions, and we found no evidence of an extended atmosphere in either of the planets. Theoretical calculations of the mass-loss rate supported our results, predicting escape rates of 4.3$\times$10$^{11}$\,g\,s$^{-1}$, 7.2$\times$10$^{12}$\,g\,s$^{-1}$ and 3.3$\times$10$^{10}$\,g\,s$^{-1}$ for GJ~9827\,b, c and d, respectively, making them unlikely to still retain their hydrogen-dominated atmosphere. 

We also made use of the {\it HST} spectra in order to characterize GJ~9827's high energy emission, which was used for the above escape rate calculations, and the ISM absorption along its sightline. We reconstructed the intrinsic \Lya\ and MgII stellar fluxes, necessary because of attenuating H I and Mg II interstellar gas between us and the star, finding $F$(\Lya) = (5.42$^{+0.96}_{-0.75}$) $\times$ 10$^{-13}$ erg cm$^{-2}$ s$^{-1}$ and $F$(MgII) = (5.64$\pm$ 0.24) $\times$ 10$^{-14}$ erg cm$^{-2}$ s$^{-1}$. We report that GJ~9827 is Doppler-shifted +30 km s$^{-1}$ from the velocity frame of the absorbing ISM gas, which results in almost negligible attenuation of the narrow Mg II lines, but dramatic absorption of the broad \Lya~line. However, the reconstructed intrinsic \Lya\ flux is inconsistent with the literature predictions based on its Mg II emission \citep{wood2005,youngblood2016}. Comparing GJ~9827 to other K dwarfs as well as M dwarfs, we found it to have a significantly high \Lya\ surface flux and a significantly low Mg II surface flux. This could have important implications on the planetary atmospheres in the system as \Lya\ and Mg II, the two brightest emission lines in GJ~9827's UV spectrum, have a large effect on atmospheric photochemistry, potentially controlling which are the dominant species in the atmosphere. GJ~9827's \Lya\ and Mg II flux discrepancy also highlights the importance of caution when using UV scaling relations for atmospheric escape calculations or photochemistry calculations. Not acknowledging the natural variability between individual stars could be detrimental to our assumptions about the composition or even presence of exoplanet atmospheres. 

%{\color{red}This could also have significant implications for our atmospheric escape calculation, where the XUV flux was estimated based on the \Lya~flux... ok getting informal here: if the \Lya~flux is anomalously high (and we're not 100\% sure if it's truly anomalous, it could be within the expected scatter), then does that mean that XUV flux is anomalously high too? Or are we overestimating it and therefore overestimating the escape rate for the planets?  (add something about being cautious about EUV scaling relations. Specific stars have specific SEDs, and not acknowledging the natural variability around these scaling relations could be detrimental to our assumptions about these exoplanets.})%UV observations play an important role in the exoplanetary field. In fact, the high-energy radiation from the star can affect the upper layers of the planetary atmosphere, inducing photochemical and escape mechanisms. Then characterizing the star becomes crucial to better characterize the planetary features. 
As a nearby system of planets transiting a bright star, GJ~9827 is being intensely studied for a variety of reasons. More {\it HST} STIS UV transit observations are planned for GJ~9827\,b, that will allow us to confirm our results presented here and investigate possible variations in the stellar flux. We have also observed transits of all three Super-Earths orbiting GJ~9827 with {\it Spitzer} (Livingston et al in prep.). These, together with our approved Cycle 1 GO CHEOPS transit observations, will further enhance dynamical constraints via transit timing variations that will provide invaluable measurements in the infrared to complement {our \textit{Hubble} observations, as well as facilitate efficient future observations (e.g., with \textit{JWST}) for a system that will be intensely characterized in the years to come.

\acknowledgments
Based on observations made with the NASA/ESA Hubble Space Telescope,
obtained from the data archive at the Space Telescope Science Institute. STScI
is operated by the Association of Universities for Research in Astronomy, Inc.
under NASA contract NAS 5-26555.

This work is partly financed by the Spanish Ministry of Economics and Competitiveness through project ESP2016-80435-C2-2-R and PGC2018-098153-B-C31.
This work is partly supported by JSPS KAKENHI Grant Numbers JP18H01265 and JP18H05439, and JST PRESTO Grant Number JPMJPR1775.

\end{document}